\documentclass[twoside,11pt]{article}

\usepackage{jmlr2e}
\usepackage{amsmath}
\usepackage{url}
\usepackage{amssymb}
\usepackage{graphicx}
\usepackage{appendix}
\usepackage{array}
\usepackage{hhline}
\graphicspath{{./}}
\usepackage{dcolumn}

\pdfoutput=1
\usepackage{hyperref}
\hypersetup{
    colorlinks=true,
    linkcolor=black,
    citecolor=black,
    filecolor=black,
    urlcolor=black,
}
\newcommand{\eff}{{\rm eff}}

\def\WAIC{\mathrm{WAIC}}

\def\LOO{\mathrm{LOO}}

\DeclareMathOperator{\N}{N}

\DeclareMathOperator{\Var}{\mathrm{Var}}

\def\f{{f}}
\def\g{{g}}
\def\y{{y}}
\def\X{{X}}
\def\x{{x}}

\DeclareMathOperator*{\argmax}{arg\,max}

\def\app#1#2{%
  \mathrel{%
    \setbox0=\hbox{$#1\sim$}%
    \setbox2=\hbox{%
      \rlap{\hbox{$#1\propto$}}%
      \lower1.1\ht0\box0%
    }%
    \raise0.25\ht2\box2%
  }%
}

\ShortHeadings{Bayesian LOO approximations for GLVMs}{Vehtari, Mononen, Tolvanen, Sivula, and Winther}
\firstpageno{1}

 \makeatletter
 \def\@starteditor{\noindent \small {}}
 \makeatother

\begin{document}

\title{Bayesian leave-one-out cross-validation approximations for Gaussian latent variable models}
\author{\name Aki Vehtari \email aki.vehtari@aalto.fi\\
  \name Tommi Mononen \\ 
  \name Ville Tolvanen \\ 
  \name Tuomas Sivula \\ 
  \addr Helsinki Institute of Information Technology HIIT, \\
  Department of Computer Science, Aalto University\\
  P.O.Box 15400, 00076 Aalto, Finland 
  \AND 
  \name Ole Winther \email owi@imm.dtu.dk\\ 
  \addr Technical University of Denmark\\
  DK-2800 Lyngby, Denmark}

\editor{}

\maketitle
\thispagestyle{plain}

\begin{abstract}
  The future predictive performance of a Bayesian model can be
  estimated using Bayesian cross-validation. In this article, we
  consider Gaussian latent variable models where the integration over
  the latent values is approximated using the Laplace method or
  expectation propagation (EP). We study the properties of several
  Bayesian leave-one-out (LOO) cross-validation approximations that in
  most cases can be computed with a small additional cost after
  forming the posterior approximation given the full data. Our main
  objective is to assess the \emph{accuracy of the approximative LOO
    cross-validation estimators}. That is, for each method (Laplace
  and EP) we compare the approximate fast computation with the exact
  brute force LOO computation. Secondarily, we evaluate the accuracy
  of the Laplace and EP approximations themselves against a ground
  truth established through extensive Markov chain Monte Carlo
  simulation. Our empirical results show that the approach based upon
  a Gaussian approximation to the LOO marginal distribution (the
  so-called cavity distribution) gives the most accurate and reliable
  results among the fast methods.
\end{abstract}
\vspace{0.5\baselineskip}


\section{Introduction}

Bayesian cross-validation can be used to assess predictive performance.
\citet{Vehtari+Ojanen:2012} provide an extensive review of theory and
methods in Bayesian predictive performance assessment including
decision theoretical assumptions made in Bayesian cross-validation.
\citet{Gelman+Hwang+Vehtari:2014} provide further details on
theoretical properties of leave-one-out cross-validation and
information criteria, and \citet{Vehtari+Gelman+Gabry:2016} provide
practical fast computation in the case of Monte Carlo posterior inference.
In this article, we present the properties of several Bayesian
leave-one-out (LOO) cross-validation approximations for {\em Gaussian
latent variable models} (GLVM) with factorizing likelihoods. 
Integration over the latent variables is performed with either the Laplace
method or expectation propagation (EP). We show that for these methods 
leave-one-out cross-validation can be computed accurately with zero or a 
negligible additional cost after forming the full data posterior approximation.
 
Global (Gaussian) and factorizing variational approximations for latent variable inference are not considered in this paper. They have the same order computational complexity as Laplace and EP but with a larger pre-factor on the dominating ${\cal O}(n^3)$ term, where $n$ is the number of observations \citep{Nickisch+Rasmussen:2008}. EP may be expected to be the most accurate method \citep[e.g.][]{Nickisch+Rasmussen:2008,Vanhatalo+Vehtari:2010,Jylanki+Vanhatalo+Vehtari:2011,Riihimaki+Jylanki+Vehtari:2013}
and Laplace to have the smallest computational overhead. So EP and Laplace may be considered the methods of choice for accuracy and speed, respectively. We expect that our overall results and conclusions for Laplace and EP carry over to Gaussian variational. For non-GLVM models such as generalized linear and deep generative models, the (factorized) Gaussian variational approximations scale to large datasets \citep{challis2013GKL,ranganath2014black,kingma2013auto,rezende2014stochastic}. It is of interest to derive approximate LOO estimators for these models, but that is outside the scope of this paper.

We consider a prediction problem with an explanatory variable $x$ and
an outcome variable $y$. The same notation is used interchangeably for
scalar and vector-valued quantities. The observed data are denoted by
$D = \{(x_{i},y_{i})\}_{i=1}^n$ and future observations by
$(\tilde{x},\tilde{y})$.  We focus on GLVMs, where the observation
model $p(y_i|f_i,\phi)$ depends on a local latent value $f_i$ and
possibly on some global parameters $\phi$, such as the scale of the measurement error process.
Latent values $\f=(f_1,\ldots,f_n)$ have a joint Gaussian prior
$p(\f|x,\theta)$ which depends on covariates $\x$ and hyperparameters
$\theta$ (e.g., covariance function parameters for a Gaussian
process). The posterior of the latent $\f$ is then
\begin{align}
  p(\f|D,\theta,\phi) \propto p(\f|\x,\theta)\prod_{i=1}^np(y_i|f_i,\phi).
\end{align}
As a specific example we use Gaussian process (GP) models
\citep[reviewed, e.g., by][]{Rasmussen+Williams:2006}, but the methods
are applicable also for other GLVMs which have the same factorizing
form (e.g. Gaussian Markov random field models used in the R-INLA software \citep{Lindgren+Havard:2015}). 
Some of the presented methods are applicable more generally,
requiring only a factorizing likelihood with terms $p(y_i|f_i,\phi)$ and
a method to integrate over the marginal posteriors $p(f_i|D,\theta,\phi)$.
The results presented in this paper can be generalized to the cases
where a likelihood term depends upon more than one latent variable
\citep[e.g.][]{Tolvanen+Jylanki+Vehtari:2014} or the latent value
prior is non-Gaussian
\citep[e.g.][]{Seeger:2008,Hernandez-Lobato+etal:2008,Hernandez-Lobato+etal:2010}.
For clarity we restrict our treatment to the case of one likelihood
term with one latent value.

We are interested in assessing the predictive performance of our
models to report this to application experts or to perform model
selection.
For simplicity, in this paper we use only the logarithmic score, but
the methods can be also be used with application specific utilities such
as classification error. Logarithmic score is the standard
scoring rule in Bayesian cross-validation (see
\citet{Geisser+Eddy:1979}) and it has desirable properties for
scientific inference
\citep{Bernardo+Smith:1994,Gneiting+Raftery:2007}.

The predictive distribution for a future observation $\tilde{y}$ given future covariate values $\tilde{x}$ is
\begin{align}
  p(\tilde{y}|\tilde{x},D)=\int p(\tilde{y}|\tilde{f},\phi)p(\tilde{f}|\tilde{x},D,\theta) p(\phi,\theta|D) d \tilde{f} d\phi d\theta.
\end{align}
 The expected predictive performance using the log score and unknown
 true distribution of the future observation $p_t(\tilde{x},\tilde{y})$ is
 \begin{align}
   \int p_t(\tilde{x},\tilde{y}) \log p(\tilde{y}|\tilde{x},D) d\tilde{x}d\tilde{y}.
 \end{align}
This expectation can be approximated by re-using the observations and
computing the leave-one-out Bayesian cross-validation estimate
\begin{align}
  \label{eq:LOO}
  \LOO = \frac{1}{n} \sum_{i=1}^n \log {p(y_i|x_i,D_{-i})},
\end{align}
where $D_{-i}$ is all other observations except $(x_i,y_i)$.
Here we consider only cases with random $\tilde{x}$ from the same
distribution as $\x$. See \citet{Vehtari+Ojanen:2012} for discussion
of fixed, shifted, deterministic, or constrained $\tilde{x}$.

In addition to estimating the expected log predictive density, it may be
interesting to look at a single value, $\log {p(y_i|x_i,D_{-i})}$. These
terms, also called conditional predictive ordinates
($\mbox{CPO}_i$), may reveal observations which are highly influential or
not well explained by the model \citep[see, e.g.,][]{Gelfand:1996}.
The probability integral transform (PIT) values ${F(y_i|x_i,D_{-i})}$, where $F$ is the predictive CDF, can be used to assess the calibration of the predictions \citep[see, e.g.,][]{Gneiting+Balabdaoui+Raftery:2007}.

The straightforward brute force implementation of leave-one-out
cross-validation requires recomputing the posterior distribution
$n$ times. Often leave-one-out cross-validation is replaced with
$k$-fold cross-validation requiring only $k$ recomputations of the
posterior, with $k$ usually 10 or less. Although $k$-fold-CV is robust
and would often be computationally feasible, there are several fast
approximations for computing LOO with a negligible additional
computational cost after forming the posterior with the full data.

Several studies have shown that the Laplace method and EP perform well (compared
to the gold standard Markov chain Monte Carlo inference) for GLVMs with
many log-concave likelihoods
\citep[e.g.][]{Rue+Martino+Chopin:2009,Vanhatalo+Pietilainen+Vehtari:2010,Martins+Simpson+Lindgren+Havard:2013,Riihimaki+Vehtari:2014}.
EP has also been shown to be close to Markov chain Monte Carlo inference for
classification models (log-concave likelihood, but potentially highly
skewed posterior) and non-log-concave likelihoods
\citep[e.g.][]{Nickisch+Rasmussen:2008,Vanhatalo+Vehtari:2010,Jylanki+Vanhatalo+Vehtari:2011,Cseke+Heskes:2011,Riihimaki+Jylanki+Vehtari:2013,Vanhatalo+gpstuff:2013,Tolvanen+Jylanki+Vehtari:2014}. In this paper we also consider the accuracy of approximative LOO with standard Markov chain Monte Carlo 
inference for LOO as our benchmark.

In practical data analysis work, it is useful to start with fast
approximations and step by step check whether a computationally more
expensive approach can improve the predictive accuracy. We propose the following three step approach:
\begin{enumerate}
\item Find the MAP estimate $(\hat{\phi},\hat{\theta})$ using the
  Laplace method to approximately integrate over the latent values
  $f$.
\item Using $(\hat{\phi},\hat{\theta})$ obtained in the previous step,
  use EP to integrate over the latent values and check whether the
  predictive performance improves substantially compared to using the
  Laplace method (we may also re-estimate $\hat{\phi}$ and
  $\hat{\theta}$).
\item Integrate over $\phi$ and $\theta$ and check whether integration
  over the parameters improves predictive performance.
\end{enumerate}
Details of the computations involved are given in Sections~\ref{sec:glvm} and \ref{sec:loo}.
Based on these steps we can continue with the model that has the best
predictive performance or the one that makes predictions fastest, or both. Often
we also need to re-estimate models when data are updated or additional
covariates become available, and then again a fast and accurate
posterior approximation is useful. To follow the above approach, we
need accurate predictive performance estimates for the Laplace method
and EP. 

The main contributions of this paper are:
\begin{itemize}
\item A unified presentation and thorough empirical comparison of
  methods for approximate LOO for Gaussian latent variable models with
  both log-concave and non-log-concave likelihoods and MAP and hierarchical
  approaches for handling hyperparameter inference (Section~\ref{sec:loo}).
\item The main conclusion from the empirical investigation (Section~\ref{sec:results}) is the observed superior accuracy/complexity
  tradeoff of Gaussian latent cavity distribution based LOO
  estimators. Although there are more accurate non-Gaussian
  approximations of the marginal posteriors, their use does not
  translate into substantial improvements in terms of LOO
  cross-validation accuracy and also introduces considerable
  instability. Using the widely applicable information criterion (WAIC) in the computation does not provide any benefits.
\item The Laplace Gaussian cavity distribution (LA-LOO) (Section~\ref{sec:la-loo}), although mentioned by \citet{Cseke+Heskes:2011},
  has not been used previously for LOO estimation. LOO consistency of
  LA-LOO using linear response theory is proved (Appendix~\ref{app:la-lr-loo}).
\item Truncated  weights quadrature integration (Section~\ref{sec:q-loo}) inspired by truncated importance sampling
  is a novel way to stabilize the quadrature used in some LOO
  computations.
\end{itemize}

\section{Gaussian latent variable models}
\label{sec:glvm}

In this section, we briefly review the notation and methods for Gaussian
latent variable models used in the rest of the article. 
We focus on Gaussian processes \citep[see,
e.g.,][]{Rasmussen+Williams:2006}, but most of the discussion also holds
for other factorizing GLVMs.
We consider models with a Gaussian prior $p(\f|\x,\theta)$ on latent
values $\f=(f_1,\ldots,f_n)$ and factorizing likelihood
\begin{align}
  p(f|D,\theta,\phi) = \frac{1}{Z} \prod_{i=1}^n p(y_i|f_i,\phi) p(\f|\X,\theta) ,
\end{align}
where $Z$ is a normalization factor and equal to the marginal likelihood $p(\y|X,\theta,\phi)= \int \prod_{i=1}^n p(y_i|f_i,\phi) p(\f|\X,\theta) df$.
For example, in the Gaussian process framework the multivariate Gaussian prior on latent values is $p(f|\x,\theta)=\N(f|\mu_0,K)$, where 
$\mu_0$ is the prior mean and $K$ is a covariance matrix constructed by a covariance function $K_{i,j} = k(\x_i, \x_j;\theta)$, which characterizes the correlation between two points. In this paper, we assume that the prior mean $\mu_0$ is zero, but the results generalize to nonzero prior means as well.

\subsection{Gaussian observation model}
With a Gaussian observation model, 
\begin{align}
  p(y_i|f_i,\sigma^2)=\N(y_i|f_i, \sigma^2),
\end{align}
where $\phi=\sigma^2$ is the noise variance, the conditional posterior
of the latent variables is a multivariate
Gaussian
\begin{align}
  p(f|D,\theta,\phi)&=\N(f|\mu,\Sigma),\nonumber \\
  \text{where}\quad \mu&=K(K+\sigma^2I)^{-1}y\nonumber \\ 
  \text{and} \quad \Sigma&=(K^{-1}+\sigma^{-2}I)^{-1}=K-K(K+\sigma^2I)^{-1}K.
\end{align}
The marginal posterior is simply $p(f_i|D,\theta,
\sigma^2)=\N(\mu_i,\Sigma_{ii})$ and the marginal likelihood
$p(\y|X,\theta, \sigma^2)$ can be computed analytically using
properties of the multivariate Gaussian \citep[see,
e.g.,][]{Rasmussen+Williams:2006}.

\subsection{Non-Gaussian observation model}
In the case of a non-Gaussian likelihood, the conditional posterior
$p(\f|{D},\theta, \phi)$ needs to be approximated. In this paper, we
focus on expectation propagation (EP) and the Laplace method (LA),
which form a multivariate Gaussian approximation of the joint latent
posterior
\begin{align}
  q(f|D,\theta,\phi) = \frac{1}{Z}p(\f|\X,\theta) \prod_{i=1}^n \tilde{t}_i(f_i),
  \label{eq:qf}
\end{align}
where the $\tilde{t}_i$ are (unnormalized) Gaussian approximations of the
likelihood contributions. We use $q$ to denote approximative joint and
marginal distributions in general, or the specific approximation used
in each case can be inferred from the context.

\subsection{Expectation propagation} 
Expectation propagation \citep{Opper+Winther:2000,Minka:2001}
approximates independent non-Gaussian likelihood terms by
unnormalized Gaussian form site approximations (aka pseudo-observations),
\begin{equation}
p(y_i | f_i) \simeq \tilde{t}_i(f_i|\tilde{Z}_i,\tilde{\mu}_i, \tilde{\Sigma}_i) = \tilde{Z}_i \N(f_i | \tilde{\mu}_i, \tilde{\Sigma}_i),
\end{equation}
where $\tilde{Z}_i=\int p(y_i | f_i) \N(f_i | \tilde{\mu}_i, \tilde{\Sigma}_i) df_i$,  and $\tilde{\mu}_i$ and $\tilde{\Sigma}_i$ are the
parameters of the site approximations, or \emph{site parameters}.
The joint latent posterior approximation is then
\begin{align}
& p(\f | D, \phi, \theta) = \frac{1}{Z}p(\f | X, \theta) \prod_i p(y_i | f_i, \phi) \nonumber \\ 
&\quad \approx \frac{1}{Z_{\mathrm{EP}}}p(\f | X, \theta) \prod_i \tilde{t}_i(f_i) = q(\f | D, \phi, \theta),
\end{align}
where $Z$ is the normalization constant or the marginal likelihood,
$Z_{\mathrm{EP}}$ is the EP approximation to the marginal likelihood
and $q(\f | D)$ is a multivariate Gaussian posterior approximation.

EP updates the site approximations by iteratively improving accuracy of
the marginals. To update the $i$th site approximation, it is first removed
from the marginal approximation to form a cavity distribution,
\begin{align}
  \label{eq:ep-cavity}
  q_{-i}(f_i) \propto q(f_i|D)/\tilde{t}_i(f_i),
\end{align}
where the marginal $q(f_i|D)$ is obtained analytically using
properties of the multivariate Gaussian.

The cavity distribution is combined with the original likelihood term to
form a more accurate marginal distribution called the tilted distribution:
\begin{align}
  \label{eq:ep-tilted}
  q_{-i}(f_{i})p(y_i|f_i,\phi).
\end{align}
Minimization of Kullback-Leibler divergence from the tilted
distribution to the marginal approximation corresponds to matching the
moments of the distributions.  
Hence for Gaussian approximation, the zeroth, first and second
moments of this tilted distribution are computed, for example, using
one-dimensional numerical integration. The site parameters are updated
so that moments of the marginal approximation $q(f_i|D)$ match the
moments of the tilted distribution $q_{-i}(f_{i})p(y_i|f_i,\phi)$. The new
$q(\f)$ can be computed after a single site approximation has been
updated (sequential EP) or after all the site approximations have been
updated (parallel EP).

\subsection{Laplace approximation}
\label{sec:LA}
The Laplace approximation is constructed from the second-order Taylor
expansion of\linebreak $\log p(\f|\y,\theta, \phi)$ around the mode $\hat{\f}$,
which gives a Gaussian approximation to the conditional posterior,
\begin{align}
  q(f|{D},\theta,\phi)=\N(f|\hat{f},\hat{\Sigma}) \approx
  p(\f|{D},\theta,\phi),
\end{align}
where $\hat{\Sigma}= (K^{-1} + \tilde{\Sigma}^{-1})^{-1}$ is the
inverse of the Hessian at the mode with $\tilde{\Sigma}$ being a
diagonal matrix with elements
\citep[e.g.,][]{Rasmussen+Williams:2006,Gelman+etal+BDA3:2013},
\begin{align}\label{eq:Sigma}
\tilde{\Sigma}_i&=-\frac{1}{\nabla_i\nabla_i \log p(y_i|f_i,\phi)|_{f_i=\hat{f}_i}}.
\end{align} 
From this joint Gaussian approximation we can analytically compute an
approximation of the marginal posterior $p(f_i|{D},\theta,\phi)$ and
the marginal likelihood $p(\y|\x,\theta,\phi)$.
The Laplace approximation can also be written as
\begin{align}
  q(f|D,\theta,\phi) = \frac{1}{Z}p(\f|\X,\theta) \prod_{i=1}^n \tilde{t}_i(f_i),
\end{align}
where $\tilde{t}_i(f_i)$ are Gaussian terms
$\N(f_i|\tilde{\mu}_i,\tilde{\Sigma}_i)$ 
with
\begin{align}\label{eq:mu}
  \tilde{\mu}_i&=\hat{f}+\tilde{\Sigma}_i\nabla_i \log p(y_i|\f_i,\phi)|_{f_i=\hat{f}_i}.
\end{align}

\subsection{Marginal posterior approximations}
\label{sec:marginal-improvements}

Many leave-one-out approximation methods require explicit computation
of full posterior marginal approximations.  We thus review alternative
Gaussian and non-Gaussian approximations of the marginal posteriors
$p(f_i|D,\theta,\phi)$ following the article by
\citet{Cseke+Heskes:2011}.
The exact joint posterior can be written as (dropping $\theta$, $\phi$ and $D$ for brevity)
\begin{align}
  p(f) \propto q(f)\prod_{i}\epsilon_i(f_i) \quad \text{with} \quad \epsilon_i(f_i) = p(y_i|\f_i,\phi)/\tilde{t}_i(\f_i),
\end{align}
where $\epsilon_i(f_i)$ is the ratio of the exact likelihood and the site
term approximating the likelihood.  By integrating over the other
latent variables, the marginal posterior can be written as
\begin{align}
   p(f_i) \propto q(f_i)\epsilon_i(f_i)\underbrace{\int q(f_{-i}|f_i)\prod_{j\neq i}\epsilon_j(f_j) df_{-i}}_{c_i(f_i)},
  \label{eq:exact_marginal}
\end{align}
where $f_{-i}$ represents all other latent variables except $f_i$. Local
methods use $\epsilon_i(f_i)$ which depends locally only on
$f_i$. Global methods additionally use an approximation of $c_i(f_i)$
which depends globally on all latent variables.
Next we briefly review different
marginal posterior approximations of this exact marginal (see Table
\ref{tab:marginal_improvements} for a summary).
\begin{table}
  \centering
  \small
  {
  \renewcommand{\arraystretch}{1.3}
  \begin{tabular}{l | c  p{8.5cm}}
    Method & Improvement & Explanation \\
    \hline
    LA-G & - & Gaussian marginal $q(f_i)$ from the joint distribution \\
    LA-L & local & tilted distribution $q(f_{i})\tilde{t}_i(\f_i)^{-1}p(y_i|f_i,\phi)$ \\
    LA-TK  & global & $q(f_{i})\tilde{t}_i(\f_i)^{-1}p(y_i|f_i,\phi)c_i(f_i)$, where $c_i(f_i )$ is approximated using the Laplace approximation \\
    LA-CM/CM2/FACT & global & $q(f_{i})\tilde{t}_i(\f_i)^{-1}p(y_i|f_i,\phi)c_i(f_i)$, where $c_i(f_i )$ is approximated using the Laplace approximation with simplifications \\
    \hline
    EP-G & - & Gaussian marginal $q(f_i)$ from the joint distribution \\
    EP-L & local & tilted distribution $q_{-i}(f_{i})p(y_i|f_i,\phi)$, where $q_{-i}(f_{i})$ is obtained as a part of EP method \\
    EP-FULL & global & $q_{-i}(f_{i})p(y_i|f_i,\phi)c_i(f_i)$, where $c_i(f_i )$ is approximated using EP \\
    EP-1STEP/FACT & global & $q_{-i}(f_{i})p(y_i|f_i,\phi)c_i(f_i)$, where $c_i(f_i )$ is approximated using EP  with simplifications\\
   \end{tabular}
 }
  \caption{Summary of the methods for computing approximate marginal posteriors. In global methods $c_i(f_i )=\int q(f_{-i}|f_i)\prod_{j\neq i}\epsilon_j(f_j) df_{-i}$ is a multivariate integral and $ \epsilon_j(f_j) = p(y_j|\f_j,\phi)/\tilde{t}_j(\f_j)$.}
  \label{tab:marginal_improvements}
\end{table}

\paragraph{Gaussian approximations.}
The simplest approximation is to use the Gaussian marginals $q(f_i)$,
which are easily obtained from the joint Gaussian obtained by the
Laplace approximation or expectation propagation; we call these LA-G and EP-G. By denoting the mean
and variance of the pseudo observations (defined by the site terms) by
$\tilde{\mu}_i$ and $\tilde{\sigma}^2_i$ respectively, the joint
approximation has the same form as in the Gaussian case:
\begin{align}
  q(f|D,\theta,\phi)&=\N(\mu,\Sigma)\nonumber\\
  \text{with}\quad \mu&=\Sigma\tilde{\Sigma}^{-1}\tilde{\mu}, \quad \text{and} \quad \Sigma=(K^{-1}+\tilde{\Sigma}^{-1})^{-1},
  \label{eq:local-global-terms}
\end{align}
where $\tilde{\Sigma}$ is diagonal matrix with
$\tilde{\Sigma}_{ii}=\tilde{\sigma}^2$. Then the marginal is simply
$q(f_i)=\N(\mu_i,\Sigma_{ii})$. 

\paragraph{Non-Gaussian approximations using a local correction.}
The simplest improvement to Gaussian marginals is to include the local
term $\epsilon_i(f_i)$, and assume that the global term
$c_i(f_i)\approx 1$.  For EP the result is the tilted distribution
$q(f_i)\epsilon_i(f_i)=q_{-i}(f_{i})p(y_i|f_i,\phi)$ which is obtained
as a part of the EP algorithm \citep{Opper+Winther:2000}. As only 
the local terms are used to compute the improvement,
\citet{Cseke+Heskes:2011} refer to it as the local improvement and denote the
locally improved EP marginal as EP-L.

For the Laplace approximation, \citet{Cseke+Heskes:2011} propose
a similar local improvement LA-L which can be written as
$q(f_{i})\tilde{t}_i(\f_i)^{-1}p(y_i|f_i,\phi)$, where the site
approximation $\tilde{t}_i(\f_i)$ is based on the second order
approximation of $\log p(y_i|f_i,\phi)$ (see Section~\ref{sec:LA}).
In Section~\ref{sec:la-loo}, we propose an alternative way to compute
the equivalent marginal improvement using a tilted distribution
$q_{-i}(f_{i})p(y_i|f_i,\phi)$, where the cavity distribution
$q_{-i}(f_{i})$ is based on a leave-one-out formula derived using
linear response theory (Appendix~\ref{app:la-lr-loo}).
The local methods EP-L and LA-L can improve the marginal posterior
approximation only at the observed $x$, and the marginal posterior at
new $\tilde{x}$ is the usual Gaussian predictive distribution.

\paragraph{Non-Gaussian approximations using a global correction.}
Global approximations also take into account the global term
$c_i(f_i)$ by approximating the multidimensional integral in Equation~\eqref{eq:exact_marginal}, again using Laplace or EP. To obtain an
approximation for the marginal distribution, the integral $c_i(f_i)$
has to be evaluated with several $f_i$ values and the computations
can be time consuming unless some simplifications are used.  Global
methods can be used to obtain an improved non-Gaussian posterior marginal
approximation also at the not yet observed $\tilde{x}$.

Using the Laplace approximation to evaluate $c_i(f_i)$ corresponds to
an approach proposed by \citet{Tierney+Kadane:1986}, and so we label the marginal
improvement as LA-TK.
\citet{Rue+Martino+Chopin:2009} proposed an approach that can be seen
as a compromise between the computationally intensive LA-TK and the local
approximation LA-L. Instead of finding the mode for each $f_i$, they
evaluate the Taylor expansion around the conditional mean obtained
from the joint approximation $q(\f)$. The method is referred to as
LA-CM.
\citet{Cseke+Heskes:2011} propose the improvement LA-CM2 which adds
a correction to take into account that the Taylor expansion is not done
at the mode.
To further reduce the computational effort,
\citet{Rue+Martino+Chopin:2009} propose additional approximations with
performance somewhere between LA-CM and
LA-L. \citet{Rue+Martino+Chopin:2009} also discuss computationally
efficient schemes for selecting values of $f_i$ and interpolation or
parametric model fitting to estimate the marginal density for other values
of $f_i$.
\citet{Cseke+Heskes:2011} propose similar approaches for EP, with
EP-FULL corresponding to LA-TK, and EP-1STEP corresponding to
LA-CM/LA-CM2. \citet{Cseke+Heskes:2011} also propose EP-FACT and
LA-FACT which use factorized approximation to speed up the computation
of the normalization terms.

The local improvements EP-L and LA-L are obtained practically for free and
all global approximations are significantly slower. See
Appendix~\ref{app:computational-complexities} for the computational
complexities of the global approximations.
Based on the results by \citet{Cseke+Heskes:2011}, EP-L is inferior to 
global approximations, but the difference is often small, and
LA-L is often worse than the global approximations. Also based on the
results by \citet{Cseke+Heskes:2011} and our own experiments, we 
chose to use LA-CM2 and EP-FACT as the global corrections in the
experiments.

\subsection{Integration over the parameters}
To marginalize out the parameters $\theta$ and $\phi$ from the previously
mentioned conditional posteriors, we can use the exact or approximated
marginal likelihood $p(\y|\x,\theta,\phi)$ to form the marginal
posterior for the parameters
\begin{align}
  p(\theta,\phi|D)\propto p(\y|X,\theta,\phi)p(\theta,\phi),
\end{align}
and use numerical integration to integrate over $\theta$ and
$\phi$. Commonly used methods include various Monte Carlo algorithms
\citep[see list of references in][]{Vanhatalo+gpstuff:2013} as well as
deterministic procedures, such as the central composite design (CCD)
method by \citet{Rue+Martino+Chopin:2009}. Using stochastic or
deterministic samples, the marginal posterior can be
approximated as
\begin{align}
p(\f|\mathcal{D}) \approx \sum_{s=1}^S p(\f|\mathcal{D}, \phi^s, \theta^s)
 w^s,
\label{eq:weighted_samples}
\end{align}
where $w^s$ is a weight for the sample $(\phi^s, \theta^s)$. 

If the marginal posterior distribution $p(\theta,\phi|D)$ is narrow,
which can happen if $n$ is large and the dimensionality of
$(\theta,\phi)$ is small, then the effect of the integration over the
parameters may be negligible and we can use Type II MAP, that is,
choose $(\hat{\phi},\hat{\theta})=\argmax_{\phi,\theta}p(\phi,\theta|D)$.

\section{Leave-one-out cross-validation approximations}
\label{sec:loo}

We start by reviewing the generic exact LOO equations, which are then
used to provide a unifying view of the different approximations in the
subsequent sections. We first review some special cases and then more
generic approximations. The LOO approximations and their abbreviations are
listed in Table~\ref{tab:LOO_approximations}. The computational
complexities of the LOO approximations have been collected in
Appendix~\ref{app:computational-complexities}.
\begin{table}
  \centering
  \small
  {
  \renewcommand{\arraystretch}{1.3}
  \begin{tabular}{l | p{12.5cm} }
    Method & Based on \\
    \hline
    IS-LOO & importance sampling / importance weighting, Section~\ref{sec:is-loo} \\
    Q-LOO & quadrature integration, Section~\ref{sec:q-loo} \\
    TQ-LOO & truncated quadrature integration, Section~\ref{sec:q-loo}  \\
    LA-LOO & same as Q-LOO with LA-L, Section~\ref{sec:la-loo} \\
    EP-LOO & same as Q-LOO with EP-L, obtained as byproduct of EP, Section~\ref{sec:ep-loo} \\
    $\WAIC_G$ & matches the first two terms of the Taylor series expansion of LOO, Section~\ref{sec:waic} \\
    $\WAIC_V$ & matches the first three terms of the Taylor series expansion of LOO, Section~\ref{sec:waic}
   \end{tabular}
 }
  \caption{Summary of the leave-one-out (LOO) cross-validation approximations reviewed.}
  \label{tab:LOO_approximations}
\end{table}

\subsection{LOO from the full posterior}
\label{sec:post-loo}

Consider the case where we have not yet seen the $i$th observation.
Then using Bayes' rule we can add information from the $i$th observation:
\begin{align}
  p(f_i|D)=\frac{p(y_i|f_i)p(f_i|x_i,D_{-i})}{p(y_i|x_i,D_{-i})},
\end{align}
again dropping $\phi$ and $\theta$ for brevity. 
Correspondingly we can remove the effect of the $i$th observation from the
full posterior:
\begin{align}
  p(f_i|x_i,D_{-i})=\frac{p(f_i|D)p(y_i|x_i,D_{-i})}{p(y_i|f_i)}
\end{align}
If we now integrate both sides over $f_i$ and rearrange the terms we get 
\begin{align}
  \label{eq:loo_from_full_posterior}
  p(y_i|x_i,D_{-i})=1/{\int\frac{p(f_i|D)}{p(y_i|f_i)}df_i}.
\end{align}
In theory this gives the exact LOO result, but in practice we usually
need to approximate $p(f_i|D)$ and the integral over $f_i$.
In the following sections we first discuss the hierarchical approach, then
the analytic, Monte Carlo, quadrature, WAIC, and Taylor series approaches
for computing the conditional version of Equation~\eqref{eq:loo_from_full_posterior}. We then consider how the different marginal
posterior approximations affect the result.

In some cases, we can compute $p(f_i|x_i,D_{-i})$ exactly or
approximate it efficiently and then we can compute the LOO predictive density for $y_i$,
\begin{align}
  p(y_i|x_i,D_{-i}) = \int p(y_i|f_i) p(f_i|x_i,D_{-i}) df_i.
\end{align}
Or, if we are interested in the predictive distribution for a new observation $\tilde{y}_i$, we can compute
\begin{align}
  p(\tilde{y}_i|x_i,D_{-i}) = \int p(\tilde{y}_i|f_i)  p(f_i|x_i,D_{-i}) df_i ,
\end{align}
which is evaluated with different values of $\tilde{y}_i$ as it is not fixed like $y_i$.

\subsection{Hierarchical approximations}
\label{sec:hier-loo}

Instead of approximating the leave-one-out predictive density
$p(y_i|x_i,D_{-i})$ directly, for hierarchical models such as GLVMs it
is often easier to first compute the leave-one-out predictive
density conditional on the parameters
$p(y_i|x_i,D_{-i},\theta,\phi)$, then compute the leave-one-out
posteriors for the parameters $p(\theta,\phi|D_{-i})$ and combine the
results
\begin{align}
  p(y_i|x_i,D_{-i})=\int p(y_i|x_i,D_{-i},\theta,\phi)p(\theta,\phi|D_{-i}) d\theta d\phi.
\end{align}
Sometimes the leave-one-out posterior of the hyperparameters is close to
the full posterior, that is, $p(\theta,\phi|D_{-i})\approx
p(\theta,\phi|D)$. The joint leave-one-out posterior can be then
approximated as
\begin{equation}
p(f_i|x_i,D_{-i}) \approx \int p(f_i|x_i,D_{-i},\theta,\phi)p(\theta,\phi|D) d\theta d\phi
\label{eq:ghosting}
\end{equation}
\citep[see, e.g.,][]{Marshall+Spiegelhalter:2003}.  This approximation is a reasonable alternative if removing
$(x_{i},y_i)$ has only a small impact on $p(\theta,\phi | D)$ but a
larger impact on $p(f_i| D, \phi, \theta)$.
Furthermore, if the posterior $p(\theta,\phi|D)$ is narrow, a Type II
MAP point estimate of the parameters $\hat{\phi},\hat{\theta}$ may
produce similar predictions as integrating over the parameters,
\begin{equation}
p(f_i|x_i,D_{-i}) \approx p(f_i|x_i,D_{-i},\hat{\theta},\hat{\phi}).
\label{eq:map-prediction}
\end{equation}

\subsection{LOO with Gaussian likelihood}
\label{sec:g-loo}

If both $p(y_i|f_i,\phi)$ and $p(f|\theta)$ are Gaussian, then we can
compute $p(f_i|x_i,D_{-i})$ analytically. Starting from the marginal
posterior we can remove the contribution of the $i$th factor in the likelihood:
\begin{align}
  p(f_i|x_i,D_{-i},\theta,\phi)&\propto\frac{p(f_i|D,\theta)}{p(y_i|f_i,\phi)}\nonumber\\
  &=\N(f_i|\mu_{-i},v_{-i}),
\end{align}
where 
\begin{align}
  \mu_{-i} & = v_{-i}(\Sigma_{ii}^{-1}\mu_i-\sigma^{-2}y_i) \nonumber\\
  v_{-i}&=\left(\Sigma_{ii}^{-1}-\sigma^{-2}\right)^{-1}.
  \label{eq:cavity}
\end{align}
These equations correspond to the cavity distribution equations in EP.

\citet{Sundararajan+Keerthi:2001a} derived the leave-one-out
predictive distribution\linebreak $p(y_i|x_i,D_{-i})$ directly from the joint
posterior using prediction equations and properties of partitioned
matrices. This gives a numerically alternative but mathematically equivalent way to compute the
leave-one-out posterior mean and variance:
\begin{align}
  \mu_{-i} & = y_i-\bar{c}_{ii}^{-1}g_i \nonumber \\
  v_{-i}&=\bar{c}_{ii}^{-1}-\sigma^2,
\end{align}
where
\begin{align}
  g_i&=\left[(K+\sigma^2I)^{-1}y\right]_i \nonumber \\
  \bar{c}_{ii}&=\left[(K+\sigma^2I)^{-1}\right]_{ii}.
\end{align}
\citet{Sundararajan+Keerthi:2001a} also provided the equation for the LOO log predictive density
 \begin{equation}
  \label{eq:gp-loo-density}
  \log p(y_i|x_i,D_{-i},\theta,\phi)
  = - \frac{1}{2}\log(2\pi) 
  - \frac{1}{2}\log \bar{c}_{ii} 
  - \frac{1}{2}\frac{g_i^2}{\bar{c}_{ii}}.
\end{equation}
Instead of integrating over the parameters,
\citet{Sundararajan+Keerthi:2001a} used the result (and its gradient)
to find a point estimate for the parameters maximizing the LOO log
predictive density.

\subsection{LOO with expectation propagation}
\label{sec:ep-loo}

In EP, the leave-one-out
marginal posterior of the latent variable is computed explicitly as a part of the
algorithm. The cavity distribution \eqref{eq:ep-cavity} is formed from
the marginal posterior approximation by removing the site
approximation (pseudo observation) using \eqref{eq:cavity} and can be
used to approximate the LOO posterior
\begin{align}
  p(f_{i}|x_i,D_{-i},\theta,\phi) \approx q_{-i}(f_i). 
\end{align}
The approximation for the LOO predictive density
\begin{align}
  \label{eq:ep-loo-lpd}
  p(y_i|x_i,D_{-i},\theta,\phi) \approx \int p(y_i|f_i) q_{-i}(f_i) df_i
\end{align}
is the same as the zeroth moment of the tilted distribution. Hence we
obtain an approximation for $p(f_{i}|x_i,D_{-i},\theta,\phi)$ and
$p(y_i|x_i,D_{-i},\theta,\phi)$ as a free by-product of the EP
algorithm. We denote this approach as EP-LOO.  For certain likelihoods
\eqref{eq:ep-loo-lpd} can be computed analytically, but otherwise
quadrature methods with a controllable error tolerance are usually
used.

The EP algorithm uses all observations when converging to its fixed
point and thus the cavity distribution $q_{-i}(f_i)$ technically
depends on the observation $y_i$. \citet{Opper+Winther:2000}
showed using linear response theory that the cavity distribution is up
to first order leave-one-out consistent. \citet{Opper+Winther:2000}
also showed experimentally in one case that the cavity distribution
approximation is accurate. 
\citet{Cseke+Heskes:2011} did not consider LOO, but compared visually
the tilted distribution marginal approximation EP-L to many global
marginal posterior improvements. Based on these results, EP-L has some
error on the shape of the marginal approximation if there is a strong
prior correlation, but even then the zeroth moment --- the LOO
predictive density --- is accurate.
Our experiments provide much more evidence of the excellent accuracy
of the EP-LOO approximation.

\subsection{LOO with Laplace approximation}
\label{sec:la-loo}

Using linear response theory, which was used by
\citet{Opper+Winther:2000} to prove LOO consistency of EP, we also prove
the LOO consistency of Laplace approximation (derivation in
Appendix~\ref{app:la-lr-loo}).  Hence, we obtain a good approximation
for $p(f_{i}|x_i,D_{-i},\theta,\phi)$ also as a free by-product of the
Laplace method. 
Linear response theory can be used to derive two alternative ways to
compute the cavity distribution $q_{-i}(f_i)$.

The Laplace approximation can be written in terms of the Gaussian
prior times the product of (unnormalized) Gaussian form site
approximations. \citet{Cseke+Heskes:2011} define the LA-L marginal
approximation as $q(f_{i})\tilde{t}_i(\f_i)^{-1}p(y_i|f_i,\phi)$, from
which the cavity distribution, that is the leave-one-out distribution, follows as
$q_{-i}(f_i)=q(f_{i})\tilde{t}_i(\f_i)^{-1}$. It can be computed using
\eqref{eq:cavity}. We refer to this approach as LA-LOO.
The LOO predictive density can be obtained by numerical integration
\begin{align}
  \label{eq:la-lr-loo-lpd}
  p(y_i|x_i,D_{-i},\theta,\phi) \approx \int q_{-i}(f_i) p(y_i|f_i,\phi) df_i.
\end{align}
An alternative way to compute the Laplace LOO derived using linear
response theory is
\begin{equation}
  p(f_i|x_i,D_{-i},\theta,\phi) \approx \N(f_i|\hat{f}_i-v_{-i} \hat{g}_i,v_{-i}),
  \label{eq:qfDm1}
\end{equation}
where $\hat{\f}$ is the posterior mode, $\hat{\g}_i=\nabla_i\log
p(\y_i|f_i)|_{f_i=\hat{f}_i}$ is the derivative of the log likelihood at the mode, and
\begin{equation}
  v_{-i} = \left( \frac{1}{\Sigma_{ii}} - \frac{1}{\tilde{\Sigma}_i} \right)^{-1} \ .
\end{equation}
If we consider having pseudo observations with means $\hat{f}_i$ and
variances $1/\hat{h}_i$, then these resemble the exact LOO equations 
for a Gaussian likelihood given in Section~\ref{sec:g-loo}.

\subsection{Importance sampling and weighting}
\label{sec:is-loo}

A generic approach not restricted to GLVMs is based on obtaining Monte
Carlo samples $(f_i^s,\phi^s,\theta^s)$ from the full posterior
$p(f_i,\phi,\theta|D)$ and approximating
\eqref{eq:loo_from_full_posterior} as
\begin{align}
  \label{eq:is-loo-pd}
  p(y_i|x_i,D_{-i})\approx\frac{1}{\frac{1}{S}\sum_{s=1}^S\frac{1}{p(y_i|f_i^s,\phi^s)}},
\end{align}
where $\theta^s$ drops out since $y_i$ is independent of $\theta^s$ given $f_i^s$ and $\phi^s$.
This approach was first proposed by \citet{Gelfand+Dey+Chang:1992} \citep[see
also,][]{Gelfand:1996} and it corresponds to importance sampling (IS)
where the full posterior is used as the proposal distribution. We
refer to this approach as IS-LOO.

A more general importance sampling form is
\begin{align}
  \label{eq:is-loo}
  p(\tilde{y}_i|x_i,D_{-i})\approx\frac{\sum_{s=1}^Sp(\tilde{y}_i|f_i^s,\phi^s)w_i^s}{\sum_{s=1}^S w_i^s},
\end{align}
where $w_i^s$ are importance weights and
\begin{align}
  w_i^s=\frac{p(f_i^s|x_i,D_{-i})}{p(f_i^s|D)}\propto\frac{1}{p(y_i|f_i^s,\phi^s)}.
\end{align}
This form shows the
importance weights explicitly and allows the computation of other
leave-one-out quantities like the LOO predictive distribution.
If the predictive density $p(\tilde{y}_i|f_i^s,\phi^s)$ is evaluated
with the observed value $\tilde{y}_i=y_i$, Equation~\eqref{eq:is-loo}
reduces to \eqref{eq:is-loo-pd}.

The approximation \eqref{eq:is-loo-pd} has the form of the harmonic mean,
which is notoriously unstable (see, e.g., Newton \& Raftery 1994).
However the leave-one-out version is not as unstable as the harmonic mean
estimator of the marginal likelihood, which uses the harmonic mean of
$\prod_{i=1}^np(y_i|f_i^s,\phi^s)$ and corresponds to using the joint
posterior as the importance sampling proposal distribution for the
joint prior.

For the Gaussian observation model, \citet{Vehtari:2001} and \citet{Vehtari+Lampinen:2002} used exact computation for
$p(y_i|x_i,D_{-i},\theta,\phi)$ and importance sampling only for
$p(\theta,\phi|D_{-i})$. The integrated importance weights are then
\begin{align}
  w_i^s\propto\frac{1}{p(y_i|x_i,D_{-i},\theta^s,\phi^s)},
\end{align}
and the LOO predictive density is
\begin{align}
  p(y_i|x_i,D_{-i})\approx\frac{1}{\sum_{s=1}^S\frac{1}{p(y_i|x_i,D_{-i},\theta^s,\phi^s)}}.
\end{align}
The same marginalization approach can be used in the case of
non-Gaussian observation models. \citet{Held+Schrodle+Rue:2010} used the
Laplace approximation, marginal improvements, and numerical
integration to obtain an approximation for
$p(y_i|x_i,D_{-i},\theta^s,\phi^s)$ (see more in Section
\ref{sec:q-loo}). \citet{Vanhatalo+gpstuff:2013} use EP and the
Laplace method for the marginalisation in the GPstuff toolbox.
\citet{Li+Qiu+Zhang+Feng:2014} considered generic latent variable models
using Monte Carlo inference, and propose to marginalise $f_i$ by
obtaining additional Monte Carlo samples from the posterior
$p(f_i|x_i,D_{-i},\theta,\phi)$.  \citet{Li+Qiu+Zhang+Feng:2014} also
proposed the name integrated IS and provided useful results illustrating
the benefits of the marginalization. As we are focusing on EP
and Laplace approximations for the latent inference, in our experiments we use IS
only for hyperparameters.

The variance of the estimate \eqref{eq:is-loo-pd} depends on the
variance of the importance weights.  The full posterior marginal
$p(f_i^s|D)$ is likely to be narrower and have thinner tails than the
leave-one-out distribution $p(f_i^s|x_i,D_{-i})$. This may cause the
importance weights to have high or infinite variance
\citep{Peruggia:1997,Epifani+MacEachern+Peruggia:2008} as rare samples
from the low density region in the tails of $p(f_i^s|D)$ may have very
large weights.

If the variance of the importance weights is finite, the central limit
theorem holds \citep{Epifani+MacEachern+Peruggia:2008}, and the
corresponding estimates converge quickly. If the variance of the raw
importance ratios is infinite but the mean exists, the generalized
central limit theorem for stable distributions holds, and the
convergence of the estimate is slower \citep{Vehtari+Gelman:2015}.

\citet{Vehtari+Gelman+Gabry:2016} propose to use Pareto smoothed
importance sampling (PSIS) by \citet{Vehtari+Gelman:2015} for
diagnostics and to regularize the importance weights in IS-LOO. Pareto
smoothed importance sampling uses the empirical Bayes estimate of
\citet{Zhang+Stephens:2009} to fit a generalized Pareto distribution
to the tail. By examining the estimated shape parameter $\hat{k}$ of
the Pareto distribution, we are able to obtain sample based estimates
of the existence of the moments
\citep{Koopman+Shephard+Creal:2009}. When the tail of the weight
distribution is long, a direct use of importance sampling is sensitive
to one or few largest values. To stabilize the estimates,
\citet{Vehtari+Gelman+Gabry:2016} propose to replace the $M$ largest
weights by the expected values of the order statistics of the fitted
generalized Pareto distribution. \citet{Vehtari+Gelman+Gabry:2016}
also apply optional truncation for very large weights following
truncated importance sampling by \citet{Ionides:2008} to guarantee
finite and reduced variance of the estimate in all cases.  Even if the
raw importance weights do not have finite variance, the PSIS-LOO estimate
still has a finite variance, although at the cost of additional
bias. \citet{Vehtari+Gelman+Gabry:2016} demonstrate that this bias is
likely to be small when the estimated Pareto shape parameter $\hat{k}<0.7$.

If the variance of the weights is finite, then 
an effective sample size estimate can be estimated as
\begin{align}
  S_{\rm eff}=1/\sum_{s=1}^S (\tilde{w}^{s})^2, 
\end{align}
where $\tilde{w}^{s}$ are normalized weights (with a sum equal to one)
\citep{Kong+Liu+Wong:1994}. This estimate is noisy if the variance is
large, but with smaller variances it provides an easily interpretable
estimate of the efficiency of the importance sampling.

Importance weighting can also be used with deterministic evaluation
points $(\phi^s,\theta^s)$ obtained from, for example, grid or CCD by
re-weighting the weights $w^s$ in \eqref{eq:weighted_samples}; see
\citet{Held+Schrodle+Rue:2010} and \citet{Vanhatalo+gpstuff:2013}.
As the deterministic points are usually used in the low dimensional case
and the evaluation points are not far in the tails, the variance of
the observed weights is usually smaller than with Monte Carlo. If the
full posterior $p(\theta,\phi|D)$ is a poor fit to each LOO
posterior $p(\theta,\phi|D_{-i})$, then the problem remains that the tails
are not well approximated and LOO is biased towards the hierarchical
approximation \eqref{eq:ghosting} that uses the full posterior of the
parameters $p(\theta,\phi|D)$.

In the ideal case, the CCD evaluation points except the modal point
would have equal weights. The CCD approach adjusts these weights based
on the actual density and importance weighting will further adjust
them, making it possible that a small number of evaluation
points have large weights. Although the CCD evaluation points have
been chosen deterministically, we can diagnose the reliability of CCD
by investigating the distribution of the weights. If there is a small
number of CCD points, we examine the effective sample size, and in
cases where the number of points exceed 280 (which happens when there
are more than 11 parameters), we also estimate the Pareto shape
parameter $\hat{k}$.

\subsection{Quadrature LOO}
\label{sec:q-loo}

\citet{Held+Schrodle+Rue:2010} proposed to use numerical integration
to approximate 
\begin{align}
  p(y_i|x_i,D_{-i},\theta,\phi)\approx 1/{\int\frac{q(f_i|D,\theta,\phi)}{p(y_i|f_i,\phi)}df_i}.
  \label{eq:qloo}
\end{align}
We call this quadrature LOO (Q-LOO), as one-dimensional numerical
integration methods are usually called quadrature. 
Given exact $p(f_i|D,\theta,\phi)$ and accurate quadrature,
this would provide an accurate result (e.g., if the true posterior is
Gaussian, quadrature should give a result similar to the analytic
solution apart from numerical inaccuracies).
However, some error will be introduced when the latent posterior is
approximated with $q(f_i|D,\theta,\phi)$. The numerical integration of
the ratio expression may also be numerically unstable if the tail of
the likelihood term $p(y_i|f_i,\phi)$ decays faster than the tail of
the approximation $q(f_i|D,\theta,\phi)$. For example, the probit
likelihood, which has a tail that goes as $\exp(-f^2/2)/f$, will be
numerically unstable if $q(f_i|D,\theta,\phi)$ is Gaussian with a
variance below one.

\citet{Held+Schrodle+Rue:2010} tested the Gaussian marginal approximation
(LA-G) and two non-Gaussian improved marginal approximations (LA-CM
and simplified LA-CM, see Section
\ref{sec:marginal-improvements}). All had problems with the tails,
although less so with the more accurate
approximations. \citeauthor{Held+Schrodle+Rue:2010} proposed to rerun
the failed LOO cases with actual removal of the data. As
\citeauthor{Held+Schrodle+Rue:2010} had 13 to 56 failures in their
experiments, the proposed approach would make LOO relatively expensive. In our
experiments with Gaussian marginal approximations LA-G/EP-G, we also
had several severe failures with some data sets. However with the non-Gaussian
approximations LA-CM2/EP-FACT, we did not observe severe failures (see
Section~\ref{sec:results}).

If we use marginal approximations EP-L or LA-L based on the tilted
distribution $q_{-i}(f_i) p(y_i|f_i,\phi)$ (see Table
\ref{tab:marginal_improvements}), we can see that the tail problem
vanishes. Inserting the normalized tilted distribution from 
\eqref{eq:qloo}, the equation reduces to
\begin{align}
  p(y_i|x_i,D_{-i},\theta,\phi) \approx \int q_{-i}(f_i) p(y_i|f_i,\phi) df_i,
\end{align}
which is the EP-LOO or LA-LOO predictive density estimate depending
on which approximation is used.

We also present an alternative form of \eqref{eq:qloo}, which gives
additional insight about the numerical stability when the global
marginal improvements are used. As discussed in Section
\ref{sec:marginal-improvements}, we can write the marginal
approximation with a global improvement as
\begin{align}
  \frac{Z_q}{Z_p}q(f_i)\tilde{t}(f_i)^{-1}p(y_i|f_i,\phi)c_i(f_i),
\end{align}
where $c_i(f_i)$ is a global correction term (see Eq.~\eqref{eq:exact_marginal}). Replacing $q(f_i)\tilde{t}(f_i)^{-1}$ with
the cavity distribution from EP-L or LA-L gives
\begin{align}
  \frac{Z_q}{Z_p}q_{-i}(f_i)p(y_i|f_i,\phi)c_i(f_i),
\end{align}
which we can insert into \eqref{eq:qloo} to obtain
\begin{align}
  p(y_i|x_i,D_{-i},\theta,\phi) \approx \frac{\int p(y_i|f_i,\phi)  q_{-i}(f_i) c_i(f_i) df_i}{\int q_{-i}(f_i) c_i(f_i) df_i}.
  \label{eq:qloo2}
\end{align}
Here $q_{-i}(f_i) c_i(f_i)$ is a global corrected leave-one-out
posterior, and we can see that the stability will depend on
$c_i(f_i)$.
The correction term $c_i(f_i)$ may have increasing tails, which is usually
not a problem in $q_{-i}(f_i) p(y_i|f_i,\phi) c_i(f_i)$, but may be a
problem in $q_{-i}(f_i) c_i(f_i)$.  In addition, the evaluation of
$c_i(f_i)$ at a small number of points and using interpolation for the
quadrature \citep[as proposed by][]{Rue+Martino+Chopin:2009} is
sometimes unstable, which may increase the instability of $\int
q_{-i}(f_i) c_i(f_i) df_i$.
Depending on the details of the computation, \eqref{eq:qloo}
and \eqref{eq:qloo2} can produce the same result up to numerical
accuracy, if the relevant terms cancel out numerically in
Equation~\eqref{eq:qloo}. This happens in our implementation with global
marginal posterior improvements, and thus in  Section~\ref{sec:results} we do
not report the results separately for \eqref{eq:qloo} and
\eqref{eq:qloo2}.

\citet{Held+Schrodle+Rue:2010} and \citet{Vanhatalo+gpstuff:2013} use
quadrature LOO in a hierarchical approximation, where the parameter
level is handled using importance weighting (Section
\ref{sec:is-loo}). Our experiments also use this
approach. Alternatively, we could approximate by integrating over the
parameters in the marginal and likelihood separately and approximate
LOO as
\begin{align}
  p(y_i|x_i,D_{-i})\approx 1/{\int\frac{q(f_i|D)}{p(y_i|f_i,D)} df_i}.
\end{align}
If the
integration over $\theta$ and $\phi$ is made using Monte Carlo or
deterministic sampling (e.g. CCD), then this is equivalent to using
quadrature for conditional terms and importance weighting of the
parameter samples.

\paragraph{Truncated weights quadrature.}

As the quadrature approach may also be applied beyond simple
GLVMs, we propose an approach for stabilizing the general form.
Inspired by truncated importance sampling by \citet{Ionides:2008}, we
propose a modification of the quadrature approach, which makes it more
robust to approximation errors in tails:
\begin{align}
  p(y_i|x_i,D_{-i},\theta,\phi)\approx\frac{\int p(y_i|f_i,\phi)\ p(f_i|D,\theta,\phi)\tilde{w}(f_i)df_i}{\int p(f_i|D,\theta,\phi)\tilde{w}(f_i)df_i},
\end{align}
where 
\begin{align}
  \tilde{w}(f_i)=\frac{1}{\max(p(y_i|f_i,\phi),c)},
\end{align}
and $c$ is a small positive constant. When $c=0$, we get the original
equation. When $c$ is larger than the maximum value of
$p(y_i|f_i,\phi)$, we get the posterior predictive density $p(y_i|D)$.

With larger values of $p(y_i|f_i,\phi)$ and $c$ we
avoid the possibility that the ratio explodes. In easy cases, where
the numerator gets close to zero before $c$ is used, we get a negligible
bias. In difficult cases, we have a bias towards the full posterior
predictive density.

 In truncated importance sampling, the truncation level is based on the
 average raw weight size and the number of samples \citep[see details in][]{Ionides:2008}.
 Following this idea we choose
$$
c^{-1}=c_0^{-1}\int_a^b \frac{p(f_i|D,\theta,\phi)}{p(y_i|f_i,\phi)} df_i.
$$ 
By limiting the integral to interval $(a,b)$, we avoid tail problems while
capturing information about the average level of the
weights. Based on experiments not reported here, we choose
$c_0=10^{-4}$ and the interval $(a,b)$ to extend 6 standard deviations from
the mode of the marginal posterior in each direction. A case-specific
$c_0$ could further improve results, but a fixed $c_0$ already shows the
usefulness of the truncation.
We refer to truncated weights quadrature LOO by TQ-LOO.  In the
experiments we show that TQ-LOO can provide more stable results than
Q-LOO. 

\subsection{Widely applicable information criterion}
\label{sec:waic}

\citet{Watanabe:2010a,Watanabe:2010d} showed that the widely applicable
information criterion (WAIC) is asymptotically equivalent to Bayesian
LOO.
\citet{Watanabe:2010a,Watanabe:2010d} provided two forms for WAIC,
which we refer to as $\WAIC_G$ and $\WAIC_V$ following
\citet{Vehtari+Ojanen:2012}. 
WAIC was originally defined on the scale of
mean negative log density, but for better cohesion within this
paper we use the scale of mean log density.
In the following discussion we drop the dependence on $\phi$ and $\theta$ and return to this point towards the end of the section.
Both WAIC forms consist of the mean training log predictive density
$\frac{1}{n} \sum_{i=1}^n \log p(y_i|D)$ and a second term to correct for its
optimistic bias. These correction terms may be
interpreted as the complexity of the model or the effective number of
parameters in the model, but the interpretation does not always seem to be clear.

The correction term in $\WAIC_G$ is based on the difference between
the training utility and Gibbs utility ($\frac{1}{n} \sum_{i=1}^n\int\log p(y_i|f_i) p(f_i|D)d f_i$)
giving
\begin{align}
  \WAIC_G = \frac{1}{n} \sum_{i=1}^n \log p(y_i|D) - 2\sum_{i=1}^n \left[ \,  \log E_{f_i|D}[p(y_i|f_i)] - E_{f_i|D}[ \log p(y_i|f_i)] \, \right] \ , 
\end{align}
where the Gibbs utility differs from the mean training log predictive
density by the changed order of the logarithm and the expectation over the
posterior.

The correction term in $\WAIC_V$ is based on the functional variance
which describes the fluctuation of the posterior distribution:
\begin{align}
  \WAIC_V = \frac{1}{n} \sum_{i=1}^n \log p(y_i|D) - \frac{1}{n} \sum_{i=1}^n \Var_{f_i|D}[\log p(y_i|f_i)].
\end{align}
Both of these criteria are easy to compute using Monte Carlo samples from
the joint posterior $p(\f|D)$, or marginal posterior approximation of
$p(f_i|D)$ and quadrature integration.

\citet{Watanabe:2010d} used a Taylor series expansion to prove the
asymptotic equivalence to Bayesian LOO with error term $O_p(n^{-2})$.
To examine this relation we write the LOO log predictive density
using condensed notation for \eqref{eq:loo_from_full_posterior}
\begin{align}
  -\frac{1}{n} \sum_{i=1}^n \log E_{f_i|D}[p(y_i|f_i)^{-1}].
\end{align}
By defining a generating function of functional cumulants,
\begin{align}
F(\alpha)=\frac{1}{n}\sum_{i=1}^n\log E_{f_i|D}[p(y_i|f_i)^\alpha],
\end{align}
and applying a Taylor expansion of $F(\alpha)$ around 0 with
$\alpha=-1$, we obtain an expansion of the leave-one-out predictive density:
\begin{align}
\LOO=F'(0)-\frac{1}{2}F''(0)+\frac{1}{6}F^{(3)}(0)-\sum_{i=4}^\infty\frac{(-1)^iF^{(i)}(0)}{i!}.
\end{align}
From the definition of $F(\alpha)$ we get
\begin{align}
  F(0)&=0\nonumber \\
  F(1)&=\frac{1}{n}\sum_{i=1}^n\log E_{f_i|D}[p(y_i|f_i)]\nonumber \\
  F'(0)&=\frac{1}{n}\sum_{i=1}^n E_{f_i|D}[ \log p(y_i|f_i)]\nonumber \\
  F''(0)&=\frac{1}{n}\sum_{i=1}^n \Var_{f_i|D}[\log p(y_i|f_i)].
\end{align}
Furthermore, the expansion for the mean training log predictive density is
\begin{align}
  F(1)=F'(0)+\frac{1}{2}F''(0)+\frac{1}{6}F^{(3)}(0)+\sum_{i=4}^\infty\frac{F^{(i)}(0)}{i!},
\end{align}
the expansion for $\WAIC_G$ is
\begin{align}
\text{WAIC}_G(n)&=F(1)-2[F(1)-F'(0)] = -F(1)+2F'(0)\nonumber\\
&=F'(0)-\frac{1}{2}F''(0)-\frac{1}{6}F^{(3)}(0)-\sum_{i=4}^\infty\frac{F^{(i)}(0)}{i!},
\end{align}
and the expansion for $\WAIC_V$ is
\begin{align}
\text{WAIC}_V(n)&=F(1)-F''(0)\nonumber\\
&=F'(0)-\frac{1}{2}F''(0)+\frac{1}{6}F^{(3)}(0)+\sum_{i=4}^\infty\frac{F^{(i)}(0)}{i!}.
\end{align}
The first two terms of the expansion of $\WAIC_G$ and
the first three terms of the expansion of $\WAIC_V$ match with the
expansion of LOO. Based on the expansion we may assume that $\WAIC_V$
is the more accurate approximation for LOO. 

\citet{Watanabe:2010d} shows that the error of $\WAIC_V$ is
$O_p(n^{-2})$ and argues that asymptotically further terms have
negligible contribution. However, the error can be significant in
the case of finite $n$ and weak prior information, as shown by
\citet{Gelman+Hwang+Vehtari:2014}, and for hierarchical models,
as demonstrated by \citet{Vehtari+Gelman+Gabry:2016}.
For example, with Gaussian processes, if $x_i$ is far from all other $x_j$, then $f_i$ has
a low correlation with any other $f_j$ and the effective number of
observations affecting the posterior of $f_i$ is close to 1.
In such cases, the higher order terms of the expansion are
significant. The higher order terms of $\WAIC_V$ match the higher
order terms of the mean training log predictive density and thus
$\WAIC_V$ will be biased towards that. This is also evident from our 
experiments (see Section~\ref{sec:results}). It is not as clear what
happens with $\WAIC_G$, but experimentally the behavior is similar but
with higher variance than with $\WAIC_V$. The performance of both
WAICs clearly also depend on the accuracy of the marginal approximation
$q(f_i|D)$.

Instead of WAIC, we could directly compute a desired number of terms
from the series expansion of LOO. In theory, we could approximate the
exact result with a desired accuracy if enough higher order functional
cumulants exist. This does not always work (e.g., if the
posterior is Cauchy and the observation model is Gaussian), but it is
true with a Gaussian prior on latent variables and a log-concave
likelihood \citep{An:1998}. In practice, the accuracy is limited by
the computational precision of the higher cumulants, which is limited
by the number of Monte Carlo samples or by the distributional
approximation $q(f_i|D)$. If the cumulants are computed using
$q(f_i|D)$ and quadrature, then the approximation based on Taylor series
expansion converges eventually to Q-LOO (within numerical accuracy).

In the above equations we had dropped dependency on $\phi$ and
$\theta$. Like in other LOO-CV approximations, the parameter level can
be handled using importance weighting. Alternatively we can handle the
parameter level in full WAIC style by computing the cumulants of the
marginal posteriors, where $\phi$ and $\theta$ have been integrated out,
and using these cumulants to compute WAIC.

WAIC is related to the deviance information criterion (DIC). We do not
review DIC here and instead refer to \citet{Gelman+Hwang+Vehtari:2014}
for the reasons we prefer WAIC to DIC. 
Indeed, in our experiments not reported here, DIC had larger error than WAIC.

\section{Results}
\label{sec:results}

Using several real data sets we present results illustrating the
properties of the reviewed LOO-CV approximations. Table
\ref{tab:datasets} lists the basic properties of four classification
data sets (Ripley, Australian, Ionosphere, Sonar), one survival data
set with censoring (Leukemia), and one data set for a Student's $t$ regression (Boston).
All data sets are available from the internet.
Several classification data sets were selected as the posterior is
likely to be skewed and there are often differences in performance
between Laplace approximation and expectation
propagation. The classification data sets have different numbers of
covariates so we can investigate to what degree this affects the accuracy of 
the LOO-CV approximations. The leukemia survival data set was selected as we
often analyze survival data with censoring.
The Boston data set for a regression with a Student's $t$ observation model
was selected to illustrate the performance in the case of a
non-log-concave likelihood, which may produce multimodal latent
posterior.
Similar results were obtained with other data sets not reported here.
\begin{table}[tp]
\centering
\begin{tabular}{ l | c c c  l}
Data set   & n    & d  &  \#($\phi,\theta$) & observation model \\  \hline
Ripley & 250 & 2 & 5 & probit \\
Australian & 690 & 14 & 17 & probit \\ 
Ionosphere & 351 & 33 & 4 & probit \\
Sonar & 208 & 60 & 4 & probit \\
Leukemia  & 1043 & 4 & 7 & log-logistic with censoring \\
Boston  & 506 & 13 & 17 & Student's $t$
\end{tabular}
\caption{Summary of data sets and models in our examples.}
\label{tab:datasets}
\end{table}

For all data sets we fit Gaussian processes with constant, linear, and
squared exponential covariance functions. When using the squared
exponential covariance function, we use a separate length scale for
each covariate except with the Ionosphere and Sonar data sets, where
we use one common length scale. For the classification data sets we
use a Bernoulli observation model with probit link. For the Leukemia
data set we use a log-logistic model with censoring \citep[as
in][p. 511]{Gelman+etal+BDA3:2013}.  For the Boston data set we use a Student's
$t$ observation model with $\nu=4$ degrees of freedom. A fixed $\nu$ was
chosen as the Laplace approximation
\citep{Vanhatalo+jylanki+vehtari:2009} had occasional problems when
integrating over an unknown $\nu$. Robust-EP by
\citet{Jylanki+Vanhatalo+Vehtari:2011} works well also with $\nu$ unknown. 
All the experiments were done using GPstuff
toolbox\footnote{Available at
  \url{http://research.cs.aalto.fi/pml/software/gpstuff/}}
\citep{Vanhatalo+gpstuff:2013}.  The Laplace method is implemented as
described in \citet{Vanhatalo+Pietilainen+Vehtari:2010}. The
Laplace-EM method for Student's $t$ model is implemented as described
in \citet{Vanhatalo+jylanki+vehtari:2009}. Parallel EP for other data
sets than Boston and parallel robust-EP for Student's $t$ models are
implemented as described in
\citet{Jylanki+Vanhatalo+Vehtari:2011}. CCD is implemented as
described in \citet{Vanhatalo+Pietilainen+Vehtari:2010}. Markov chain
Monte Carlo (MCMC) sampling is based on elliptical slice sampling for
latent values \citep{Murray+Adams+MacKay:2010} and surrogate slice
sampling \citep{Murray+Adams:2010} for jointly sampling latent values
and hyperparameters.
The practical speed comparisons of the posterior and LOO approximation
methods are shown in Appendix~\ref{app:practical-speed}.

Although in the review we described the estimation of the expected
performance $\LOO = \frac{1}{n} \sum_{i=1}^n \log
{p(y_i|x_i,D_{-i})}$, below we report $n\times\LOO$. 
For these data sets this puts the approximation errors
for all sets on the same scale.
This scale has two other interpretations. First, the difference
between the sum training log predictive density and $n\times\LOO$ can be
interpreted sometimes as the effective number of parameters measuring
the model complexity \citep{Vehtari+Ojanen:2012,Gelman+Hwang+Vehtari:2014}. Second, the
significance of the difference between two models can be approximately
calibrated if $n\times\LOO$ is interpreted as a pseudo log Bayes factor and if a
similar calibration scale is used as for the Bayes factor
\citep{Vehtari+Ojanen:2012}. As a rule of thumb, based upon asymptotic theory and experience we would like the
approximation error for $n\LOO$ to be smaller than 1. See the additional
discussion of using Bayesian cross-validation in model selection in
\citet{Vehtari+Lampinen:2002} and \citet{Vehtari+Ojanen:2012}. We let $\LOO_i\equiv \log
{p(y_i|x_i,D_{-i})}$ and $\widehat{\LOO}_i$ be the corresponding approximate quantity. In the tables we report a bias and deviation of individual terms as 
\begin{align}
{\rm Bias} & = \sum_{i=1}^n (\widehat{\LOO}_i-\LOO_i)  \\ 
{\rm Std}^2 & = \sum_{i=1}^n (\widehat{\LOO}_i-\LOO_i-{\rm Bias})^2 .
\end{align}  
The acronyms used in the following are MCMC=Markov chain
Monte Carlo, CCD=central composite design, MAP=Type II maximum a
posteriori, PSIS=Pareto smoothed
importance sampling, and those listed in Tables~\ref{tab:marginal_improvements}
and \ref{tab:LOO_approximations}.

\subsection{Exact LOO comparison to MCMC}

The ground truth exact LOO results were obtained by brute force
computation of each $p(y_i|x_i,D_{-i})$ separately by leaving out the
$i$th observation. We do that for each method: Laplace, EP and
MCMC. MCMC serves as the golden standard for the posterior inference
to which we compare Laplace and EP.
We show results separately for estimating the predictive performance
with and without a global correction (CM2/FACT). As discussed in Section
\ref{sec:marginal-improvements}, only the global corrections produce
non-Gaussian predictive distributions for the latent variable
$\tilde{f}$ at a new point $\tilde{x}$.
Our main interest is in approximating $p(y_i|x_i,D_{-i})$, but we also show
exact LOO results for the conditional $p(y_i|x_i,D_{-i},\phi,\theta)$
with fixed parameters $\theta,\phi$, which were obtained by optimizing
the marginal posterior $p(\theta,\phi|D)$ (type II MAP).  In this
case, LOO-CV is unbiased only conditionally as it does not take into
account the effect of the fitting of the parameters
$\theta,\phi$. However, it is useful to first evaluate the accuracy of
approximations for $p(y_i|x_i,D_{-i},\theta,\phi)$, as these can be
used with integrated importance sampling (see
Section~\ref{sec:is-loo}) for hierarchical computation of
$p(y_i|x_i,D_{-i})$.

The first part of Table~\ref{tab:MCMC-vs} shows the exact LOO
results with hyperparameters fixed to Laplace Type II MAP. LA has
similar performance to MCMC for all data sets except Ionosphere and
Sonar, for which LA is significantly inferior. LA-CM2 is able to improve
the predictive performance for the Sonar dataset to be similar with
MCMC, and for the Ionosphere, the performance is even better than for
MCMC.

The second part of Table~\ref{tab:MCMC-vs} shows
the exact LOO results with hyperparameters fixed to EP Type II MAP. EP
has similar performance to MCMC for all data sets and EP-FACT is not
able improve the performance. The small differences between MCMC
results conditional on either LA-MAP or EP-MAP fixed hyperparameters are due to
differences in the marginal likelihood approximations of LA and EP
leading to different MAP estimates. However, this difference between
LA-MAP and EP-MAP results is less interesting than differences with
full integration.

The third part of Table~\ref{tab:MCMC-vs} shows the exact LOO
results with hyperparameters integrated with MCMC or CCD. LA+CCD is as
good as MCMC for the Ripley, Australian and Leukemia data sets. LA-CM2+CCD
improves the predictive performance for Ionosphere and Sonar. The
performance of LA-CM2+CCD for Sonar is even better than MCMC and
EP(-FACT)+CCD. LA-CM2+CCD failed to produce an answer in about 9\% of leave-one-out
rounds (the LA-CM2 method
failing with some hyperparameter values) and thus no result is shown. EP is as good as MCMC for all
data sets other than Boston and EP+FACT is not able to improve the performance
at all.

Overall, when integrating over the hyperparameters, the difference
between the predictive performance of LA and EP is small except for the
Ionosphere and Sonar data sets. LA(-CM2)+CCD and EP(-FACT)+CCD have
significantly worse predictive performance than MCMC for the 
Student's $t$ regression with the Boston data. Since LA(-CM2) and EP(-FACT) performed as well as
MCMC with fixed hyperparameters, the worse performance of CCD is due
to error in the approximation of the marginal likelihood
\citep[see][]{Jylanki+Vanhatalo+Vehtari:2011} and full MCMC is able to
find better hyperparameters during the joint sampling of the latent
values and hyperparameters.

\begin{table}[tp]
\centering
\footnotesize
\begin{tabular}{ l  l | l  l  l  l  l  l}
 & Method      & Ripley  & Australian & Ionosphere & Sonar      & Leukemia     & Boston \\ \hline
\multicolumn{2}{l|}{$\theta,\phi$ fixed to LA-MAP} & & & & & & \\
  & MCMC        &{\bf -68}      &{\bf -217}      &{\bf -54}      &{\bf -68}      &{\bf -1627}      &{\bf -1097}      \\
  & LA          &     -70  (0.6)&{\bf -220} (2.8)&     -72  (2.4)&     -77  (1.7)&{\bf -1626} (0.3)&{\bf -1098} (3.0)\\
  & LA-CM2      &{\bf -68} (0.1)&{\bf -217} (0.6)&     -49  (0.7)&{\bf -67} (0.9)&{\bf -1626} (0.2)&{\bf -1098} (2.8)\\ 
\hline %  & & & & & & & \\
\multicolumn{2}{l|}{$\theta,\phi$ fixed to EP-MAP} & & & & & & \\
  & MCMC        &{\bf -69}      &{\bf -211}      &{\bf -54}      &{\bf -64}      &{\bf -1626}      &{\bf -1098}       \\
  & EP          &{\bf -68} (0.1)&{\bf -211} (0.5)&{\bf -54} (0.3)&{\bf -64} (0.2)&{\bf -1627} (0.3)&{\bf -1095}  (3.3)\\
  & EP-FACT     &{\bf -68} (0.1)&{\bf -211} (0.4)&{\bf -54} (0.3)&{\bf -64} (0.2)&{\bf -1627} (0.3)&{\bf -1094}  (3.2)\\
\hline %  & & & & & & & \\
\multicolumn{2}{l|}{$\theta,\phi$ integrated} & & & & & & \\
  & MCMC        &{\bf -70}      &{\bf -228}      &{\bf -56}      &{\bf -66 }     &{\bf -1631 }     &{\bf -1063 }      \\
  & LA+CCD      &{\bf -71} (0.5)&{\bf -230} (2.7)&     -74  (2.9)&     -79  (1.4)&{\bf -1631} (0.5)&     -1116 (6.3)  \\   
  & LA-CM2+CCD  &{\bf -69} (0.2)&{\bf -228} (1.2)&     -51  (1.5)&     -69  (1.6)&{\bf -1631} (0.5)&        NA (NA)   \\
  & EP+CCD      &{\bf -70} (0.2)&{\bf -226} (3.0)&{\bf -57} (0.5)&{\bf -65} (0.3)&{\bf -1631} (0.5)&     -1113 (5.1)  \\
  & EP-FACT+CCD &{\bf -70} (0.2)&{\bf -226} (3.1)&{\bf -57} (0.5)&{\bf -65} (0.3)&{\bf -1631} (0.5)&     -1113 (5.1) 
\end{tabular}
\caption{Exact LOO (with brute force computation) using MCMC, Laplace (LA), Laplace with CM2 marginal corrections (LA-CM2), EP or EP with FACT marginal corrections (EP-FACT) for the latent values $f$, and fixed hyperparameters $\phi,\theta$ (type II MAP) or integration over the hyperparameters with MCMC or CCD. The values in the parentheses are standard deviations of the pairwise differences from the corresponding MCMC result. Bolded values are not significantly different from the best accuracy in the corresponding category. NA indicates failed computation.}
\label{tab:MCMC-vs}
\end{table}

\subsection{Approximate LOO comparison to exact LOO -- fixed hyperparameters}

As discussed in Section~\ref{sec:hier-loo}, we compute LOO densities
$p(y_i|x_i,D_{-i})$ hierarchically by first computing the conditional LOO
densities $p(y_i|x_i,D_{-i},\theta,\phi)$.  As the accuracy of the full
LOO densities depends crucially on the conditional LOO densities, we
first analyze the LOO approximations conditional on fixed
hyperparameters. The ground truth in this section are the LA, LA-CM2, EP,
and EP-FACT results shown in Table~\ref{tab:MCMC-vs}.

Table~\ref{tab:LA-LOO-MAP} shows results when \emph{the ground truth
  is exact LOO with fixed parameters and Laplace approximation without
  a global correction} (LA in Table~\ref{tab:MCMC-vs}). LA-LOO
gives the best accuracy for all data sets by a significant margin.  Quadrature
LOO with Gaussian approximation of the latent marginals (Q-LOO-LA-G)
produces bad results for the classification data sets and
sometimes completely fails. The posterior marginals in the case of the
Leukemia model are so close to Gaussian that Q-LOO-LA-G also provides
a useful result. Truncated quadrature (TQ-LA-LOO-G) is more stable,
but it cannot fix the whole problem. Using more accurate marginal
approximation improves WAICs. $\WAIC_V$ with the LA-L marginal
approximation gives useful results for the two simplest data sets.
\begin{table}[tp]
\centering
\footnotesize
\begin{tabular}{ l | c  c  c  c  c  c}
Method         & Ripley           &  Australian     & Ionosphere       & Sonar            & Leukemia           & Boston \\ \hline
LA-LOO         & {\bf 0.01} (0.02)& {\bf 0.1} (0.04)& {\bf -0.2} (0.05)& {\bf -0.2} (0.03)& {\bf -0.0} (0.001) & {\bf -2.7} (1.0) \\ 
Q-LOO-LA-G     &     -1379   (431)&       NA    (NA)&        NA    (NA)&     -6732   (747)&      0.38  (0.02)  &      79  (6) \\     
TQ-LOO-LA-G    &      -1.2   (0.4)&      -10     (1)&      -5.6   (2.4)&       -22     (3)&       2.0  (0.3)   &      87  (6) \\     
$\WAIC_G$-LA-G &      -1.5   (1.1)&       11     (2)&       -81    (10)&       -11     (3)&       1.2  (0.05)  &      101  (7) \\     
$\WAIC_V$-LA-G &      -8.5   (6.6)&     -9.4   (5.4)&      -616    (91)&       -75    (11)&      0.40  (0.02)  &      81  (7) \\      
$\WAIC_G$-LA-L &       0.8   (0.2)&       21     (2)&        23     (2)&        26     (2)&       0.8  (0.04)  &      54  (4) \\      
$\WAIC_V$-LA-L &       0.3   (0.1)&      6.9   (0.8)&        16     (2)&        17     (2)&      0.02  (0.002) &      15  (3)     
\end{tabular}
\caption{Bias and standard deviation when the ground truth is exact LOO with Laplace and fixed full posterior MAP hyperparameters (LA in Table~\ref{tab:MCMC-vs}). Bolded values have significantly smaller absolute value than the values from the other methods for the same data set. NA indicates that computation failed.}
\label{tab:LA-LOO-MAP}
\end{table}

Table~\ref{tab:EP-LOO-MAP} shows results when \emph{the ground truth
  is exact LOO with fixed parameters and expectation propagation
  without a global correction} (EP in Table~\ref{tab:MCMC-vs}). EP-LOO
gives the best accuracy for all data sets by a significant margin. Other results
are similar to the Laplace case, that is, all methods except EP-LOO
fail badly for several data sets. Only the Ripley and Leukemia data
sets are easy enough for most of the methods to produce useful
accuracy.
\begin{table}[tp]
\centering
\footnotesize
\begin{tabular}{l | c  c  c  c  c  c }
Method         & Ripley          &  Australian    & Ionosphere     & Sonar            & Leukemia   & Boston \\ \hline
EP-LOO         & {\bf 0.2} (0.1) & {\bf 1.6} (0.5)& {\bf 0.3} (0.4)& {\bf -0.5} (0.1) & {\bf -0.0} (0.003) & {\bf -1.1} (0.9)\\
Q-LOO-EP-G     &     -352  (171) &       NA  (NA) &       NA  (NA) &        NA   (NA) &      0.02  (0.003) &      33  (3)\\
TQ-LOO-EP-G    & {   -0.2} (0.2) & {     14} (8)  & {     20} (4)  & {      NA}  (NA) & {     1.7} (0.4)   & {    44} (4)\\
$\WAIC_G$-EP-G &      0.7  (0.2) &       59  (8)  & {\bf 0.5} (3)  &       -42   (4)  &       0.8  (0.04)  &      76  (5)\\
$\WAIC_V$-EP-G &     -0.2  (0.4) &     -4.3  (7)  &      -94  (11) &      -804   (64) &      0.03  (0.004) &      37  (3)\\
$\WAIC_G$-EP-L &      0.7  (0.2) &       81  (8)  &       23  (3)  &        48   (4)  &       0.8  (0.04)  &      81  (5)\\
$\WAIC_V$-EP-L &      0.4  (0.1) &       54  (6)  &       17  (2)  &        42   (4)  &      0.02  (0.003) &      26  (3)\\
\end{tabular}
\caption{Bias and standard deviation when the ground truth is exact LOO with EP and fixed full posterior MAP hyperparameters (EP in Table~\ref{tab:MCMC-vs}). Bolded values have significantly smaller absolute values than the values from the other methods for the same data set. NA indicates that computation failed.}
\label{tab:EP-LOO-MAP}
\end{table}

Table~\ref{tab:LOO-LA-c-MAP} shows results when \emph{the ground truth
  is exact LOO with fixed parameters and Laplace approximation with
  LA-CM2 global correction} (LA-CM2 in Table
\ref{tab:MCMC-vs}). Quadrature LOO with LA-CM2 approximation of the
latent marginals (Q-LOO-LA-CM2) has the best accuracy
for all data sets except for Boston, but the
accuracy is satisfactory only for the Ripley and Leukemia datasets. Here LA-LOO has
a negative bias as the global correction
LA-CM2 can improve the marginal approximation and therefore also the
expected performance estimated with exact LOO. The results for
truncated quadrature (TQ-LOO-LA-CM2) are not reported in the table as
with adaptive truncation it produced the same results as
quadrature LOO (Q-LOO-LA-CM2). $\WAIC_V$ performs better than
$\WAIC_G$, but worse than Q-LOO-LA-CM2.
\begin{table}[tp]
\centering
\footnotesize
\begin{tabular}{l | l  l  l  l  l  l}
Method           & Ripley         &  Australian    & Ionosphere     & Sonar          & Leukemia           & Boston \\ \hline
LA-LOO           & {   -1.4} (0.6)& {   -3.3} (3.3)& {    -23}   (3)& {    -11} (2)  & {\bf 0.00} (0.1)   & {\bf -2.6} (2.2) \\ 
Q-LOO-LA-CM2     & {\bf 0.3} (0.1)& {\bf 3.1} (0.5)& {\bf 9.0} (1.8)& {\bf 7.4} (0.9)& {\bf 0.01} (0.0004)& {    11} (2) \\	    
$\WAIC_G$-LA-CM2 & {    1.0} (0.2)& {     25} (3)  & {     16}   (3)& {     27} (3)  & {    0.8} (0.04)   & {    61} (4)  \\	    
$\WAIC_V$-LA-CM2 & {    0.5} (0.1)& {     11} (2)  & {     13}   (2)& {     20} (3)  & {  0.02} (0.002) & {    22} (3)
\end{tabular}
\caption{Bias and standard deviation when the ground truth is exact LOO with Laplace-CM2 and fixed full posterior MAP hyperparameters (LA+CM2 in Table~\ref{tab:MCMC-vs}). Bolded values have significantly smaller absolute values than the values from the other methods for the same data set.}
\label{tab:LOO-LA-c-MAP}
\end{table}

Table~\ref{tab:LOO-EP-c-MAP} shows results when \emph{the ground truth
  is exact LOO with fixed parameters and expectation propagation with
  EP-FACT global correction} (EP-FACT in Table
\ref{tab:MCMC-vs}). EP-LOO provides significantly better accuracy for the
Sonar and Leukemia data sets than the other methods. 
EP-LOO also gives the best accuracy for
the other data sets, but not significantly better than 
quadrature with EP-FACT approximation of the latent marginals
(Q-LOO-EP-FACT). In addition, for the Ripley data set all methods
except $\WAIC_G$ provide good results. The EP-LOO using the EP-L tilted
distribution approximation is already good and the global correction
does not change the result much. Small errors in the quadrature
integration cumulate and Q-LOO-EP-FACT produces slightly worse results
than EP-LOO.
\begin{table}[tp]
\centering
\footnotesize
\begin{tabular}{l | l  l  l  l  l  l}
Method            & Ripley           &  Australian    & Ionosphere     & Sonar             & Leukemia & Boston \\ \hline
EP-LOO            & {\bf 0.13} (0.08)& {\bf 1.8} (0.6)& {\bf 0.1} (0.5)& {\bf -0.64} (0.08)& {\bf -0.0} (0.002) & {\bf   -1.4} (0.7)\\
Q-LOO-EP-FACT     & {\bf 0.15} (0.04)& {\bf 3.2} (0.8)& {\bf 0.9} (0.3)& {      1.1}  (0.3)& {     4.3} (1.3)   & {    4.8} (1.1) \\   
$\WAIC_G$-EP-FACT & {    0.86} (0.22)& {     82}   (8)& {     24}   (3)& {       48}    (4)& {     5.1} (1.3)   & {    81} (5)\\   
$\WAIC_V$-EP-FACT & {\bf 0.31} (0.10)& {     54}   (6)& {     17}   (2)& {       42}    (4)& {     4.4} (1.3)   & {    27} (3)
\end{tabular}
\caption{Bias and standard deviation when the ground truth is exact LOO with EP-FACT and fixed full posterior MAP hyperparameters (EP+FACT in Table~\ref{tab:MCMC-vs}). Bolded values have significantly smaller absolute values than the values from the other methods for the same data set.}\label{tab:LOO-EP-c-MAP}
\end{table}

\subsection{LOO and WAIC with varying model flexibility}
\begin{figure}[p]
   \centering
       \includegraphics[width=.9\textwidth,clip]{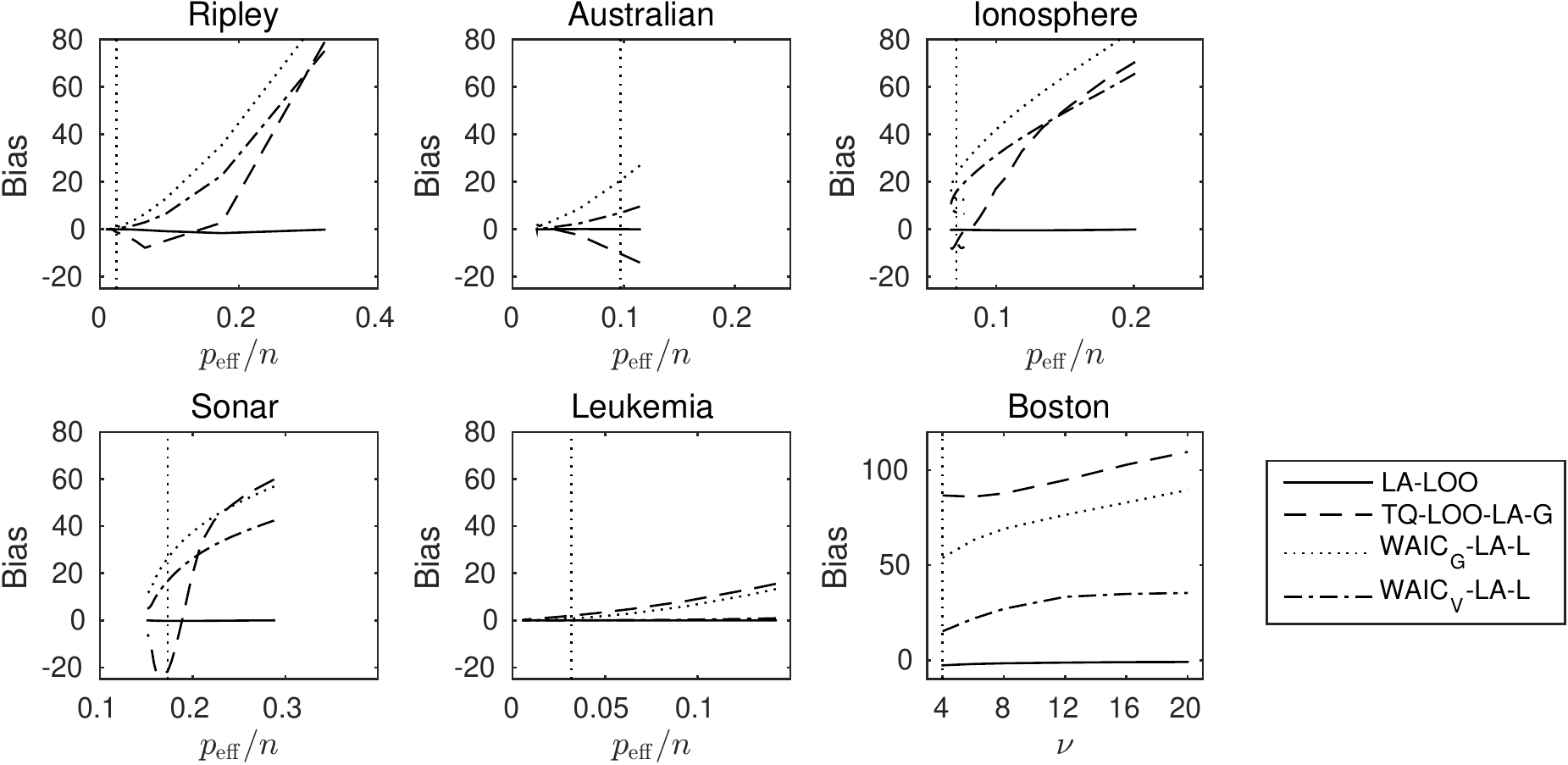}
       \caption{Bias when the ground truth is exact LOO with Laplace
         (LA in Table\ref{tab:MCMC-vs}) and varying flexibility of the
         model, or degrees of freedom in the Student's $t$ model for the Boston data. Model
         flexibility was varied by rescaling the length scale(s) in
         the GP model. Model flexibility is measured by the relative
         effective number of parameters $p_\eff/n$. The flexibility of
         the MAP model is shown with a vertical dashed line. For
         the Student's $t$ model the vertical dashed line is at $\nu=4$.}
   \label{fig:varyl_la}
 \end{figure}

\begin{figure}[p]
   \centering
       \includegraphics[width=.9\textwidth,clip]{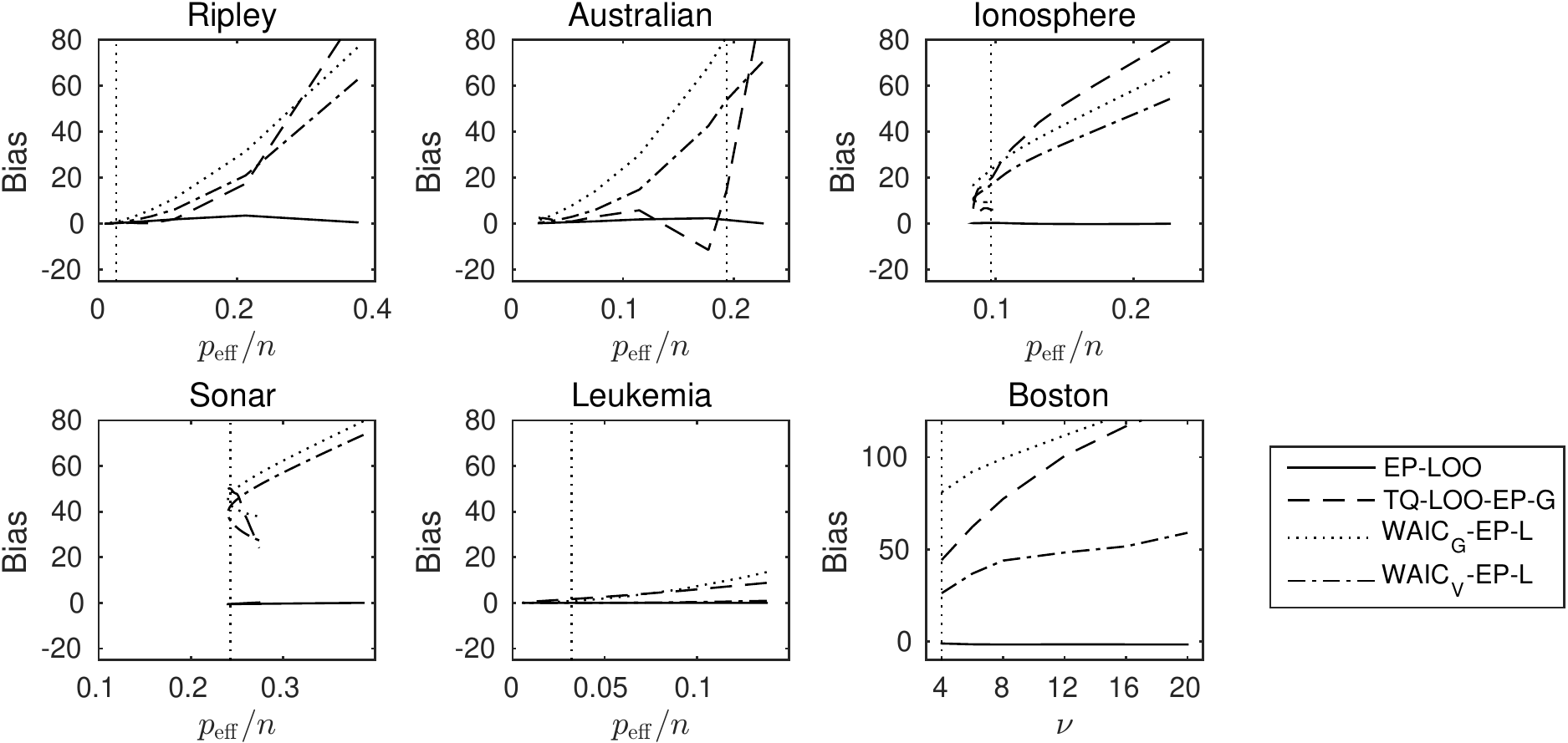}
       \caption{Bias when the ground truth is exact LOO with EP (EP in
         Table\ref{tab:MCMC-vs}) and varying flexibility of the model,
         or degrees of freedom in the Student's $t$ model for the Boston
         data. Model flexibility was varied by rescaling the length
         scale(s) in the GP model. Model flexibility is measured by
         the relative effective number of parameters $p_\eff/n$.  The
         flexibility of the MAP model is shown with a vertical dashed
         line. For the Student's $t$ the vertical dashed line is at $\nu=4$.}
   \label{fig:varyl_ep}
 \end{figure}

\begin{figure}[p]
   \centering
       \includegraphics[width=.9\textwidth,clip]{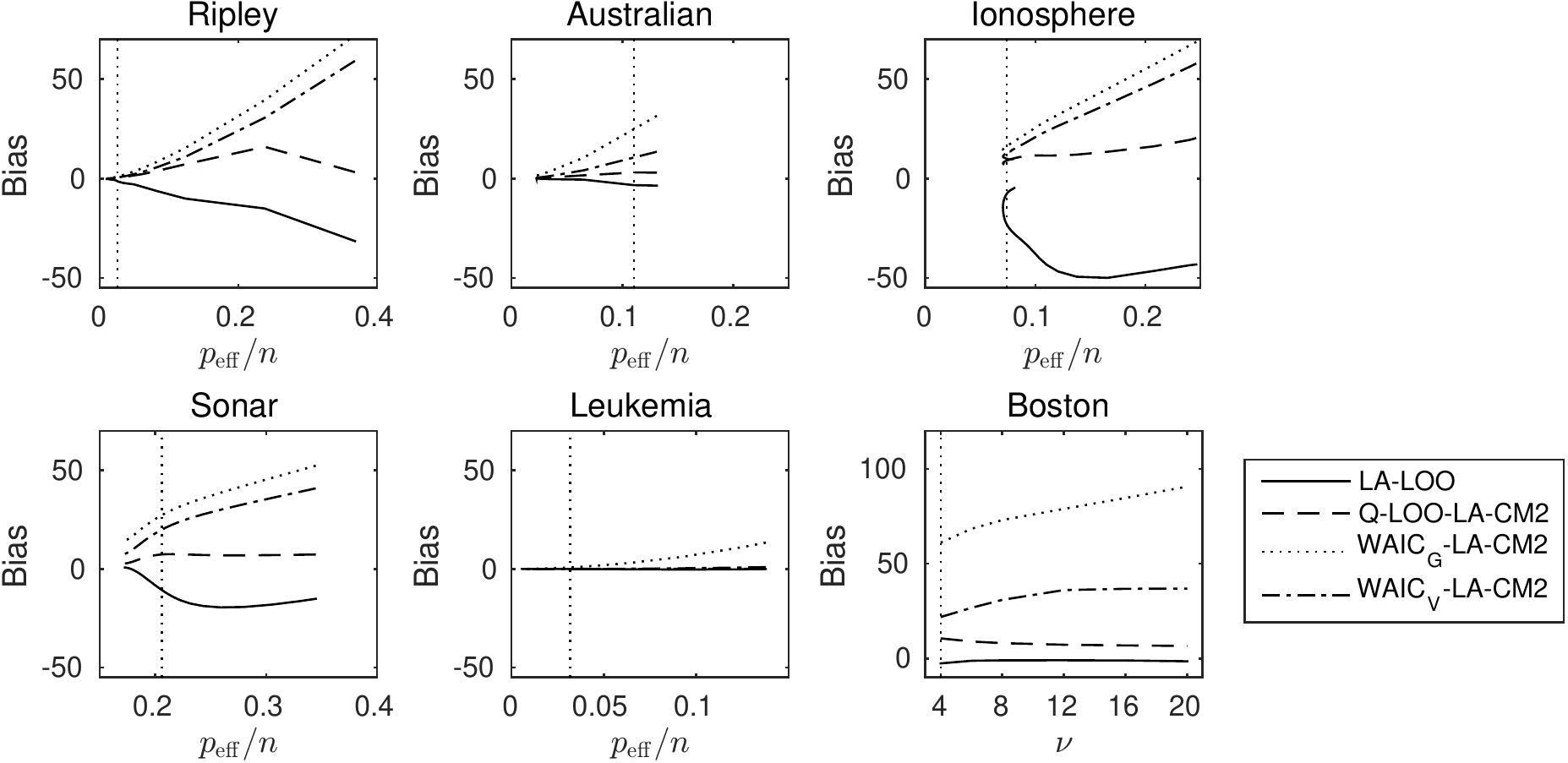}
       \caption{Bias when the ground truth is exact LOO with
         Laplace-CM2 (LA-CM2 in Table\ref{tab:MCMC-vs}) and varying
         flexibility of the model, or degrees of freedom in the Student's $t$ model for the Boston
         data. Model flexibility was varied by rescaling the length
         scale(s) in the GP model. Model flexibility is measured by
         the relative effective number of parameters $p_\eff/n$.  The
         flexibility of the MAP model is shown with a vertical dashed
         line. For the Student's $t$ the vertical dashed line is at $\nu=4$.}
   \label{fig:varyl_la_c}
 \end{figure}

\begin{figure}[p]
   \centering
       \includegraphics[width=.9\textwidth,clip]{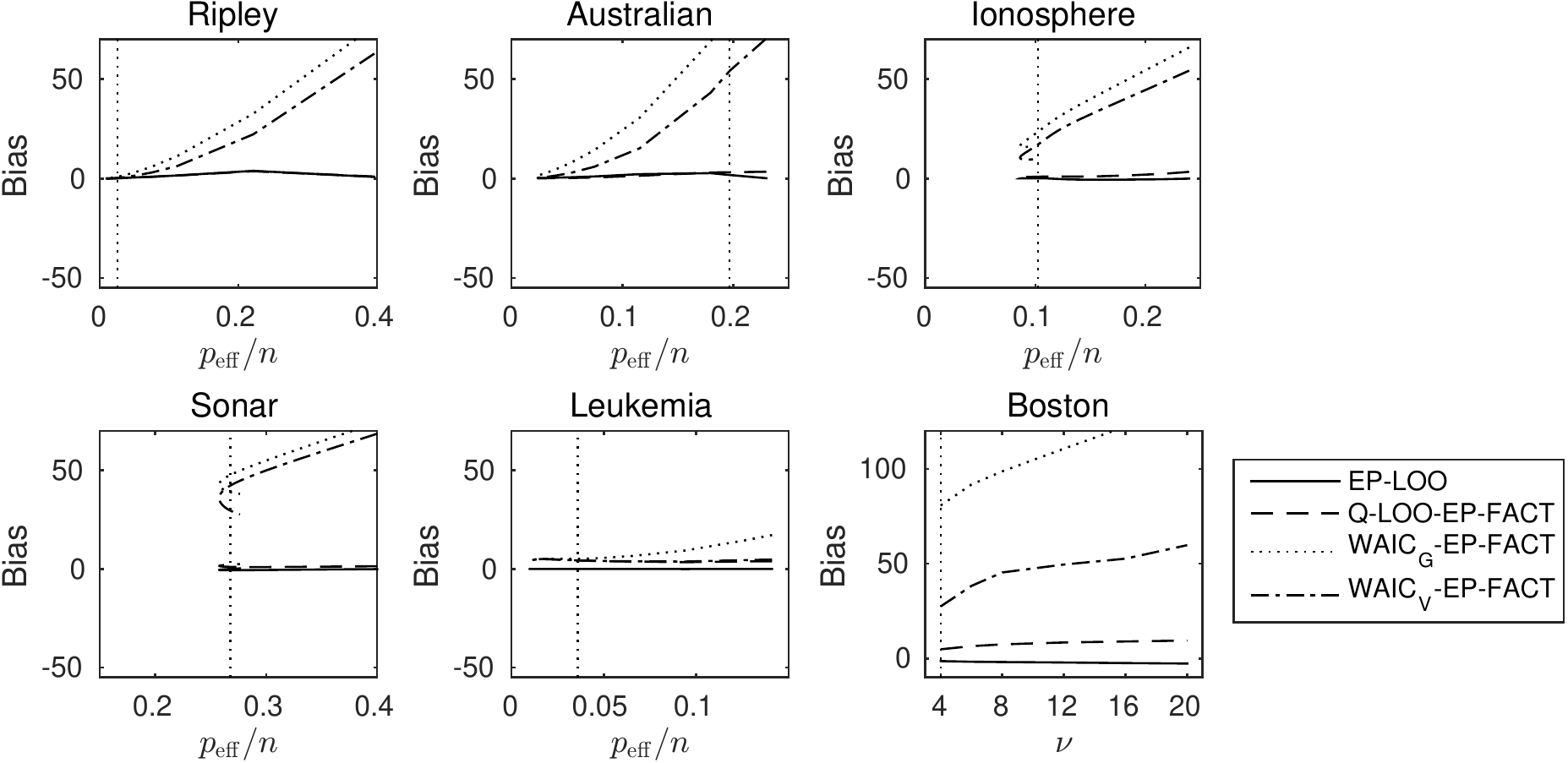}
       \caption{Bias when the ground truth is exact LOO with EP-FACT
         (EP-FACT in Table\ref{tab:MCMC-vs}) and varying flexibility
         of the model, or degrees of freedom in the Student's $t$ model for the Boston
         data. Model flexibility was varied by rescaling the length
         scale(s) in the GP model. Model flexibility is measured by
         the relative effective number of parameters $p_\eff/n$.  The
         flexibility of the MAP model is shown with a vertical dashed
         line. For the Student's $t$ the vertical dashed line is at $\nu=4$.}
   \label{fig:varyl_ep_c}
 \end{figure}

Above we saw that the methods other than LA-LOO and EP-LOO had more
difficulties with most of the data sets and especially with data sets
with a large number of covariates. Figures
\ref{fig:varyl_la}--\ref{fig:varyl_ep_c} illustrate how the
flexibility of the Gaussian process models affects the performance of
the approximations. We took the models with MAP parameter values and
re-ran the models and LOO tests, varying the length scales for all
data sets except Boston (see later). With a smaller length scale, the
GPs are more flexible and more non-linear.  With a larger length scale
GPs approach the linear model. We measure the flexibility by the
difference between the mean training log predictive density and $\LOO$,
which can be interpreted as the degree to which the model has fit to the data
or the relative effective number of parameters ($p_\eff/n$).
When the length scale gets smaller, there will be more such $f_i$s that
have a low correlation with any other $f_j$. 
In this case the full marginal posterior and LOO
marginal posterior are likely to be more different and most LOO
approximations become less accurate. This phenomenon will also occur
more easily in the case of many covariates, because more data points will tend to be
located at the corners of the data.
Figures~\ref{fig:varyl_la}--\ref{fig:varyl_ep_c} show that LA-LOO and
EP-LOO work well with different flexibilities. 
All the other methods have difficulties when the
model flexibility increases and the marginal distribution and the
cavity distribution are more different. If we look at the accuracy for
each $i$, the methods other than LA-LOO and EP-LOO start to fail when
the estimated $p_{\eff,i}$ is larger than 10\%--20\%. As a quick overall
rule of thumb,  methods other than LA-LOO and EP-LOO start to fail
when the relative effective number of parameters ($p_\eff/n$) is
larger than 2\%--5\%.

Figures~\ref{fig:varyl_la}--\ref{fig:varyl_ep_c} also show for Boston data how the degrees of freedom $\nu$ in the Student's $t$ observation model affects the accuracy. When $\nu$ increases, the observation model is closer to Gaussian and the latent posterior is more likely to be unimodal. Although the latent posterior is easier to approximate with a Gaussian when $\nu$ is large, the posterior is less robust to influential observations (``outliers'') and the error made by the methods other than LA-LOO and EP-LOO increases.

\subsection{Approximate LOO comparison to exact LOO -- hierarchical model}

Next we examine the accuracy of hierarchical LOO approximation of
$p(y_i|x_i,D_{-i})$ (see Section~\ref{sec:hier-loo}), where
the conditional LOO densities $p(y_i|x_i,D_{-i},\theta,\phi)$ are
approximated with LA-LOO or EP-LOO, which we found performed best for
conditional densities (see previous section).

Table~\ref{tab:LOO-LA-CCD} shows the results when \emph{the ground
  truth is exact LOO with CCD used to integrate over the parameter
  posterior and the Laplace method is used to integrate over the
  latent values} (LA+CCD in Table~\ref{tab:MCMC-vs}). The Laplace
approximation combined with type II MAP parameter estimates or CCD
integration but no importance weighting has an error size
related to the number of hyperparameters $(\theta,\phi)$. The unweighted
CCD or MAP gives a small error only if the number of parameters
$(\theta,\phi)$ is small.  Importance weighting of CCD works well for
all data sets except Australian and Boston. These data sets have more
parameters (17) than the others (4-8), making the inference more
difficult.  The minimum relative effective sample sizes (Ripley=60\%,
Australian=16\%, Ionosphere=59\%, Sonar=70\%, Leukemia=36\%,
Boston=0.3\%) correctly indicate that importance weighting for
Australian and Boston data sets is unreliable.

Table~\ref{tab:LOO-EP-CCD} shows the corresponding results when
\emph{the ground truth is exact LOO with CCD used to integrate over
  the parameter posterior and expectation propagation used to
  integrate over the latent values} (EP+CCD in Table
\ref{tab:MCMC-vs}). EP with the unweighted CCD or MAP gives a small
error only if the number of parameters $(\theta,\phi)$ is small.
Importance weighting of CCD works well for all data sets except
Australian and Boston.  
Again the minimum relative effective sample sizes (Ripley=60\%,
Australian=12\%, Ionosphere=36\%, Sonar=65\%, Leukemia=35\%,
Boston=9\%) correctly indicate that importance weighting for
Australian and Boston is unreliable.

As CCD integration provided good results for exact LOO (Table
\ref{tab:MCMC-vs}), the larger errors of CCD+IS for the Australian and
Boston data is not due to CCD itself failing, but importance
weighting failing. As an additional check we sampled the
hyperparameters with MCMC (6000 samples with slice sampling) and
computed Pareto smoothed importance sampling estimates (MCMC+PSIS) shown also in
Tables~\ref{tab:LOO-LA-CCD} and \ref{tab:LOO-EP-CCD}. Due to larger
number of samples, the errors are slightly reduced, but still for
the Australian and Boston data sets the errors are larger. The PSIS
diagnostics (maximum of Pareto shape parameters $\hat{k}$ for Laplace:
Ripley=$0.4$, Australian=$1.2$, Ionosphere=$0.4$, Sonar=$0.4$, Leukemia=$0.2$,
Boston=$1.3$; for EP: Ripley=$0.3$, Australian=$1.6$, Ionosphere=$0.3$,
Sonar=$0.3$, Leukemia=$0.2$, Boston=$0.7$) correctly indicate the
problematic cases ($\hat{k}>0.7$).

If the minimum relative effective sample size or PSIS diagnostics warn
about potential problems, depending on the application it may be
necessary to run, for example, $k$-fold cross-validation.

\begin{table}[tp]
\centering
\footnotesize
\begin{tabular}{ l | c  c  c  c  c  c}
  Method        & Ripley         &  Australian    & Ionosphere      & Sonar             & Leukemia & Boston \\ \hline
\it {\scriptsize LA-LOO+MCMC+PSIS}\hspace{-1mm} &  \it 0.08 (0.15)  &\it      -1.1 (2.3)&\it  0.2 (0.1)&\it  0.1 (0.2)&\it  -0.04 (0.19) &\it -6.1 (2.4)\\
LA-LOO+CCD+IS & {\bf 0.18} (0.10)& {\bf 3.4} (0.4)& {\bf -0.1} (0.1)& {\bf -0.13} (0.06)& {\bf 0.56} (0.05) & {\bf -5.2} (5.0)\\
LA-LOO+CCD    & {    0.8} (0.2)& {    7.2} (0.9)& {     0.6} (0.2)& {      0.5}  (0.2)& {    4.8}   (0.2) & {    17} (3) \\
LA-LOO+MAP    & {    1.0} (0.2)& {    9.2} (1.8)& {     1.3} (0.2)& {      1.3}  (0.3)& {    4.9}   (0.6) & {    15} (3) 
\end{tabular}
\caption{Bias and standard deviation when the ground truth is exact LOO with Laplace and CCD (LA+CCD in Table~\ref{tab:MCMC-vs}). Bolded values have significantly smaller absolute error than the values from the other methods for the same data set.}
\label{tab:LOO-LA-CCD}
\end{table}
\begin{table}[tp]
\centering
\footnotesize
\begin{tabular}{ l | c  c  c  c  c  c}
Method        & Ripley           &  Australian    & Ionosphere     & Sonar             & Leukemia & Boston \\ \hline
\it {\scriptsize EP-LOO+MCMC+PSIS}\hspace{-1mm} &\it  0.38 (0.17)&\it  -2.4 (3.4)&\it 0.8 (0.5)&\it  -0.23 (0.22)&\it  -0.16 (0.23)  &\it -0.9 (1.0)\\
EP-LOO+CCD+IS & {\bf 0.42} (0.14)& {\bf 7.3} (1.4)& {\bf 0.8} (0.6)& {\bf -0.24} (0.14)& {\bf 0.49} (0.04)  & {\bf 2.2} (1.0)\\
EP-LOO+CCD    & {     1.3}  (0.4)& {     15}   (2)& {    2.8} (1.3)& {      0.6}  (0.3)&  {    4.8}   (0.2) & {    20} (2)\\
EP-LOO+MAP    & {     1.4}  (0.3)& {     17}   (2)& {    2.8} (0.7)& {      0.9}  (0.3)&  {    4.9}   (0.6) & {    17} (2)
\end{tabular}
\caption{Bias and standard deviation when the ground truth is exact LOO with EP and CCD (EP+CCD in Table~\ref{tab:MCMC-vs}). Bolded values have significantly smaller absolute error than the values from the other methods for the same data set.}
\label{tab:LOO-EP-CCD}
\end{table}

\section{Discussion}

We have shown that LA-LOO and EP-LOO provide fast and accurate
conditional LOO results when the predictions at new points are made
using the Gaussian latent value distribution. If the predictions at new
points are made using non-Gaussian distributions obtained from the global
correction, then quadrature LOO gives useful results, but it would be
faster and more accurate to just use EP without the global
correction. Both Laplace-LOO and EP-LOO can be combined with
importance sampling or importance weighted CCD to get fast and
accurate full Bayesian leave-one-out cross-validation results.

If other methods than LA-LOO or EP-LOO are used, we propose the
following rule of thumb for diagnostics: The methods other than LA-LOO
and EP-LOO start to fail when the relative effective number of
parameters ($p_\eff/n$) is larger than 2\%--5\%.

Here we have considered fully factorizing likelihoods, but the methods
can be extended for use with likelihoods with grouped factorization,
such as in multi-class classification, multi-output regression, and
some hierarchical models with lowest level grouping. We assume that
the accuracy using Laplace-LOO and EP-LOO would also be good in these
cases.

In this paper, we have concentrated on how well exact LOO can be
estimated with fast approximations. LOO is useful for estimating the
predictive performance of of a model or in model comparison, but it should
not be used to select a single model among a large number of models
due to a selection induced bias as demonstrated by  \citet{Piironen+Vehtari:2016}.

\section*{Acknowledgments}

We thank Jonah Gabry, Andrew Gelman, Alan Saul, Arno Solin, and
anonymous reviewers for helpful comments.  We acknowledge the
computational resources provided by Aalto Science-IT project.

\begin{appendices}
  \section{Linear response Laplace leave-one-out}
  \label{app:la-lr-loo}

  Using linear response theory, used by
  \citet{Opper+Winther:2000} to prove LOO consistency of EP, we here
  derive approximative Laplace leave-one-out equations.

  The idea is to express the posterior mode solution for the LOO
  problem in terms of the solution for the full problem. The
  computationally cheap solution can be obtained by making the
  assumption that the difference between these two solutions is small
  such that their difference may be treated as a second order Taylor
  expansion. We will give two different derivations of the result
  stated in Section~\ref{sec:la-loo}; One is based on a second order
  expansion of the log likelihood and the second on a classical linear
  response argument.

  In the expansion approach we make the approximation that when
  example $i$ is removed we can treat the change in the mode for the
  remaining variables to second order. The log prior is already
  quadratic so it is only the non-linearity in the log likelihood
  terms $j\neq i$ that we expand to second order:
  \begin{equation}
    \log p(y_j|f_j,\phi) \approx \log p(y_j|\hat{f}_j,\phi) + (f_j-\hat{f}_j) \nabla_j \log p(y_j|f_j,\phi)|_{f_j=\hat{f_j}} - \frac{(f_j-\hat{f}_j)^2}{2\tilde{\Sigma}_j} \ ,
  \end{equation}
  where $\tilde{\Sigma}_j $ is defined in Equation~\eqref{eq:Sigma}. We collect the first and second order
  contributions of the expansion to give the Gaussian type leave $i$
  out factors for the likelihood terms $j\neq i$. We recognize that
  {\it these approximate factors coincide with those introduced in the
    full Laplace approximation} in Equation~\eqref{eq:qf}. We can now
  write the approximate leave one out posterior as
  \begin{equation}
    q(f|D_{-i},\theta,\phi) \propto \prod_{j\neq i} \tilde{t}_j(f_j) p(f|X,\theta)
  \end{equation}
  and the marginal as
  \begin{align}
    p(f_i|D_{-i},\theta,\phi) \approx q_{-i}(f_i) & \propto \frac{1}{\tilde{t}_i(f_i)} \int \prod_j \tilde{t}_j(f_j) p(f|X,\theta) df_{-i} \nonumber \\ 
    & \propto \frac{1}{\tilde{t}_i(f_i)}\int {\N}(f|\hat{f},\hat{\Sigma}) df_{-i} = \frac{{\N}(f_i|\hat{f}_i,\hat{\Sigma}_{ii})}{\tilde{t}_i(f_i)} \ .
  \end{align}
  This result shows that in a self-consistent second order
  approximation, where we take into account  both the explicit removal
  of likelihood term $i$ and the implicit effect on the remaining
  variables, the leave one out posterior is obtained simply by
  dividing by the Gaussian factor for $i$.
  Finally we complete the square and obtain the result in Equation~\eqref{eq:qfDm1}.

  Next we show how the same result can be obtained by a linear
  response argument. The equation for the mode is
  \begin{align}
    K^{-1} \hat{\f} = \hat{\g} \ ,
  \end{align}
  where $\hat{\g}=\nabla\log p(\y|\hat{f},\phi)$ is the vector of
  derivatives of the terms in the log likelihood (depending
  non-linearly on $\hat{f}$). Because this defines an equation for the
  mode, we only need to make a variation to first order in this case
  to recover the result we obtained above. When we remove likelihood
  term $i$ the change in the mode can be written as
  \begin{align}\label{eq:deltafhat}
    K^{-1} \delta \hat{\f} = \delta \hat{\g}
  \end{align}
  with the change in $\hat{g}$ to first order
  \begin{align}
    \delta \hat{\g} \approx -\tilde{\Sigma}^{-1} \delta \hat{f} + e_i \tilde{\Sigma}^{-1}_i \delta \hat{f}_i - e_i \hat{g}_i,
  \end{align}
  where we have used $\frac{\partial \hat{g}_i}{\partial \hat{f}_i} =
  -\hat{h}_i = -\tilde{\Sigma}^{-1}_i$ and $e_i$ is a unit vector in
  the $i$th direction.
  The first two terms on the right hand side are the indirect change
  of the equation due to the removal of term $i$ and the last is the
  direct contribution. We can now solve the linearized equation with
  respect to $\delta \hat{f}$ using the definition of the Laplace
  covariance $\Sigma = (K^{-1} + \tilde{\Sigma}^{-1} )^{-1}$
  \begin{equation}
    \delta \hat{f} = \Sigma e_i ( \tilde{\Sigma}^{-1}_i \delta \hat{f}_i - \hat{g}_i ) .
  \end{equation}
  Specializing to $\delta \hat{f}_i$ we get
  \begin{equation}
    \delta \hat{f}_i = \Sigma_{ii} ( \tilde{\Sigma}^{-1}_i \delta \hat{f}_i - \hat{g}_i )
  \end{equation}
  which can be solved with respect to $\delta \hat{f}_i$ to give $\delta
  \hat{f}_i = - v_{-i} \hat{g}_i $. This is in agreement with the change in
  the mode equation \eqref{eq:qfDm1} we found above. The variance term
  can be derived with a related linear response argument \citep{Opper+Winther:2000}.

  \section{Computational complexities}
  \label{app:computational-complexities}

  We summarize here the computational complexities of different
  methods in the paper. We first summarize the computational
  complexities of the Laplace method, expectation propagation, and
  marginal approximations used to obtain the full data posterior and
  its marginals. Then we summarize the additional computational
  complexities of the LOO methods. The related practical speed comparison
  results are shown in Appendix~\ref{app:practical-speed}.

The computational complexity for both the Laplace method and EP for
GLVMs is dominated by matrix computations related to the covariance or
precision matrix. We denote this basic cost as $c_{\rm{basic}}$. For
GLVMs with a full rank dense covariance matrix (such as Gaussian
processes used in Section~\ref{sec:results}), $c_{\rm{basic}}$ scales
with $n^3$. For reduced rank approximations in Gaussian processes such
as FITC \citep{Quinonero-Candela+Rasmussen:2005}, $c_{\rm{basic}}$
scales with $m^2n$, where $m \ll n$ is the reduced rank (affecting the
flexibility of GP). For sparse precision (in Gaussian Markov random
field models \citep[see, e.g.,][]{Rue+Martino+Chopin:2009}) or covariance matrices (in compact support covariance
function GPs \citep[see, e.g.,][]{Vanhatalo+Vehtari:2008}), $c_{\rm{basic}}$ scales with $n_{\rm{nonzeros}}^2$,
where $n < n_{\rm{nonzeros}} < n^2/2$ is the number of non-zeros in
the precision, covariance, or Cholesky matrix \citep[see more detailed
analysis of sparse GLVMs in][]{Cseke+Heskes:2011}.

For fixed $\phi$ and $\theta$, the computation of the conditional posterior and marginal likelihoods scales for
the Laplace method with $n^{\rm{Newton}}_{\rm{steps}} \times c_{\rm{basic}}$ and for EP with $n^{\rm{EP}}_{\rm{steps}} \times (c_{\rm{basic}} + n \times n_{\rm{quad}})$, 
where $n_{\rm{quad}}$ is the number of potential quadrature evaluations to compute moments (for a probit classification model the moments can be computed in closed form).
  
After the last step of the Newton or EP algorithm, the additional computational complexities for different LOO methods are shown in Table~\ref{tab:loo-comp-complex}. EP-LOO has zero additional complexity as the LOO log predictive density is computed as part of the algorithm. LA-LOO and methods using Gaussian marginals require $n$ quadrature integrals to obtain log predictive densities and thus have negligible additional complexity. EP-FACT and LA-CM2 based methods have significantly larger additional complexity. The additional complexity of EP-FACT based methods scale with $n^2 \times n_{\rm{quad,1}} \times n_{\rm{quad,2}}$, where $n_{\rm{quad,1}}$ and $n_{\rm{quad,2}}$ refer to two different quadratures in the method. The additional complexity of LA-CM based methods scale with $n \times c_{\rm{basic}} \times n_{\rm{quad}} > n^3 \times n_{\rm{quad}}$, which can be more than the complexity for the conditional posterior.
\begin{table}[tp]
\centering
\small
\begin{tabular}{ l | l }
Method         & Additional computational complexity \\ \hline
EP-LOO         &  $0$ \\
LA-LOO         &  $n \times n_{\rm{quad}}$ \\
(T)Q-LOO-LA/EP-G & $n \times n_{\rm{quad}}$ \\
$\WAIC_{G/V}$-LA/EP-G/L & $n \times n_{\rm{quad}}$ \\
(T)Q-LOO-EP-FACT & $n^2 \times n_{\rm{quad,1}} \times n_{\rm{quad,2}}$ \\
$\WAIC_{G/V}$-EP-FACT & $n^2 \times n_{\rm{quad,1}} \times n_{\rm{quad,2}}$ \\
(T)Q-LOO-LA-CM2 & $n \times c_{\rm{basic}} \times n_{\rm{quad}}$\\
$\WAIC_{G/V}$-LA-CM2 & $n \times c_{\rm{basic}} \times n_{\rm{quad}}$\\
Exact brute force LOO EP &  $n \times (n^{\rm{EP}}_{\rm{steps}} \times (c_{\rm{basic}} + n \times n_{\rm{quad}}))$ \\
Exact brute force LOO Laplace &  $n \times (n^{\rm{Newton}}_{\rm{steps}} \times c_{\rm{basic}} + n_{\rm{quad}})$
\end{tabular}
\caption{Additional computational complexity of LOO methods for fixed $\theta$ and $\phi$ after obtaining the full posterior approximation with the Laplace method or EP.}
\label{tab:loo-comp-complex}
\end{table}

The computational complexity for the Type II MAP solution is the computational complexity of forming the conditional posterior given $\theta$ and $\phi$ times the number of marginal posterior evaluations in optimisation. The additional computation to obtain LOO after Type II MAP is the computation of LOO with fixed $\theta$ and $\phi$.

The computational complexity for integration over the marginal
posterior of $\theta$ and $\phi$ is the computational complexity of
forming the conditional posterior given $\theta$ and $\phi$ times the
number of marginal posterior evaluations in the (deterministic or
stochastic) algorithm forming the posterior approximation. The
additional computation for LOO requires the computation of LOO with
fixed $\theta$ and $\phi$ for each point in the final marginal
posterior approximation and computation of importance weights which
has a negligible additional cost.

\section{Practical speed comparison}
\label{app:practical-speed}
To further give some idea of the practical speed differences between
the different algorithm implementations we show examples of
computation times for computing the marginal likelihood and LOO given
fixed $\theta$ and $\phi$. The speed comparisons were run with a
laptop (Intel Core i5-4300U CPU @ 1.90GHz x 4 + 8GB memory). As one of
the reviewers was interested in comparison to global Gaussian
variational method, we have included it in this speed comparison,
verifying the previous results that it is much slower than EP
\citep{Nickisch+Rasmussen:2008}. As GPstuff does not have the global
Gaussian variational approximation, we use the KL method in GPML
toolbox \citep{Rasmussen+Nickish-GPML:2010} to compute that. To take
into account potential general speed differences between GPstuff and
GPML we also timed GPML Laplace and EP methods. Computations were
timed several times so that caching of the previous computations were not used.

Table~\ref{tab:loo-comp-speed-Z} shows the time to compute the latent posterior and marginal
likelihood with fixed hyperparameters. In optimization (or gradient
based MCMC), the computation of gradients would have additional
computational cost. When hyperparameters are integrated out, the
approximative computation time is these multiplied by the number of
unique parameter values evaluated when obtaining the marginal
posterior samples (there are additional overheads and potential
speed-ups).
For probit, where the moments required in EP can be computed analytically,
GPstuff-EP is about 1.5--5 times slower than GPstuff-LA. For
log-logistic with censoring GPstuff-EP is about 18 times slower due to
slow quadrature based moment computations (which could be made
faster). For the Student's $t$ model GPstuff-LA and GPstuff-EP have
similar performance, as the robust Laplace-EM method by
\citet{Vanhatalo+jylanki+vehtari:2009} is slower than basic Laplace
approximation.
GPstuff-LA and GPML-LA have quite similar speed, GPstuff being
slightly faster.
GPstuff-EP is 10-25 times faster than GPML-EP, which is probably due to
using parallel updates and better vectorization allowed by parallel
updates. GPstuff has  robust-EP implementation which also works
for non-log-concave likelihoods such as Student's $t$.
Although KL has the same $O(n^3)$ computational scaling as
EP, its computational overhead makes GPML-KL 70-500 times slower than GPML-EP.
\begin{table}[tp]
\centering
\footnotesize
\begin{tabular}{ l | c c l c c c c c}
          &    &   &          &  \multicolumn{2}{c}{GPstuff}      &  \multicolumn{3}{c}{GPML} \\ 
Data set  &  n &  d&   lik    &  LA&  EP$^1$&  LA&  EP$^2$&  KL\\ \hline
Ripley    & 250&  2& probit   &      0.02  &       0.04    &   0.06  &     0.71   &   155   \\
Australian& 690& 14& probit   &      0.13  &       0.40    &   0.26  &       10   &   704   \\
Ionosphere& 351& 33& probit   &      0.05  &       0.13    &   0.08  &      1.7   &   516   \\
Sonar     & 208& 60& probit   &      0.03  &       0.04    &   0.06  &     0.47   &   233   \\
Leukemia  &1043&  4& log-logistic w. cens. &0.18 &  3.5    &   NA$^3$&     NA$^3$ &   NA$^3$ \\
Boston    & 506& 13& Student's $t$&  1.1$^4$&       1.1    &   NA$^5$ &     NA$^6$ &   39$^7$
\end{tabular}
\caption{Time (in seconds) to compute the posterior and marginal
  likelihood with fixed hyperparameters. $^1$GPstuff uses parallel EP
  \citep{Gerven+Cseke+Oostenvald+Heskes:2009} except for Student's $t$
  parallel robust-EP \citep{Jylanki+Vanhatalo+Vehtari:2011} is used.
  $^2$GPML uses random order sequential EP
  \citep{Rasmussen+Williams:2006}. $^3$GPML does not have log-logistic
  model with censoring. $^4$For Student's $t$ Laplace-EM method
  \citep{Vanhatalo+jylanki+vehtari:2009} was used. $^5$The GPML's
  Laplace inference algorithm did run without errors, but the results
  were really bad (difference in LOO-LPD $1.4e4$). $^6$GPML does not
  support EP for non-log-concave likelihoods. $^7$For Student's $t$
  the performance of global Gaussian variational (KL) method was much
  worse than the performance of Laplace-EM and EP (difference in
  LOO-LPD $147$)}
\label{tab:loo-comp-speed-Z}
\end{table}

Table~\ref{tab:loo-comp-speed-loo} shows time to compute the LOO for
fixed parameters after the full posterior has been computed (see Table~\ref{tab:loo-comp-speed-Z}).  When hyperparameters are integrated out,
the approximative computation time is these multiplied by the number
of parameter samples approximating the marginal posterior. There is no
added computational cost of going from EP to EP-LOO and the time is
spent retrieving the stored result. The computational cost of LA-LOO
is computing cavity distributions and one quadrature. Here Q-LOO
computations also include the time to compute the marginal corrections
(LA-CM2 and EP-FACT), which make them much slower.
\begin{table}[tp]
\centering
\footnotesize
\begin{tabular}{ l | c c c c c c}
          &        &        & Q-LOO- &Q-LOO- &Exact LOO& Exact LOO \\
Data set  & LA-LOO & EP-LOO & LA-CM2 &EP-FACT & Laplace & EP \\ \hline
Ripley    &   0.01 & 0.005  &     30       &  3.7         & 6.3        &   13         \\
Australian&   0.11 & 0.005  &    672       &   15         &   90       &   323        \\
Ionosphere&   0.03 & 0.005  &     91       &   6.1        &   19       &    47        \\
Sonar     &   0.02 & 0.005  &     19       &   3.0        &  7.2       &    12        \\
Leukemia  &   0.89 & 0.005  &   2547       & 11876        &  198       &  3762        \\
Boston    &   0.47 & 0.005  &    237       &   7.5        &  583       &   587
\end{tabular}
\caption{Time (in seconds) to compute LOO for fixed parameters
  after the full posterior has been computed. Here Q-LOO computations
  also include the time to compute the marginal corrections (LA-CM2 and
  EP-FACT).}
\label{tab:loo-comp-speed-loo}
\end{table}

\end{appendices}

\newpage

\bibliography{cvapprox.bib}
\end{document}